\documentclass[article]{aa} 
\usepackage{graphicx}
\usepackage{txfonts}
\usepackage[colorlinks=true,citecolor=blue]{hyperref}
\usepackage{graphicx}
\usepackage{times}
\usepackage{natbib}
\usepackage{color}
\usepackage{multirow}
\usepackage{todonotes}
\usepackage{appendix}
\usepackage{array}
\usepackage{verbatim}

\newcommand{\enzo}{\it{\small ENZO}}

\begin{document} 

\title{Cosmological simulations of the generation of cluster-scale radio emission from turbulent re-acceleration}

\author{L. Beduzzi\inst{1} \fnmsep\thanks{\email{lucabeduzzi@gmail.com}}
          \and
          F. Vazza\inst{1,2,}
          \and
           V. Cuciti\inst{1,2}
           \and
          G. Brunetti\inst{2} 
          \and
           M. Brüggen\inst{3}
           \and
          D. Wittor\inst{3}
          }
\institute{Dipartimento di Fisica e Astronomia, Università di Bologna, Via Gobetti 92/3, 40121 Bologna, Italy
         \and
            Istituto di Radio Astronomia, INAF, Via Gobetti 101, 40121 Bologna, Italy
         \and
             Hamburger Sternwarte, University of Hamburg, Gojenbergsweg 112, 21029 Hamburg, Germany
        \and Osservatorio Astronomico di Padova, INAF, Vicolo Dell'Osservatorio 5, 35122, Padova, Italy
        }

   \date{Received / Accepted}


\abstract    
{The recent discovery of so-called mega radio halos as a new class of diffuse, steep-spectrum radio sources in clusters of galaxies has raised questions about the origin and the evolution of cluster-wide radio emission.} 
{We investigate whether the formation mechanisms of radio halos and mega radio halos differ, or whether they can be produced by different modalities of the same (re-)acceleration mechanism.
Here we present results of a cosmological simulation of a disturbed galaxy cluster, with the aim to study the origin of mega radio halos.} 
{We analysed the evolution of cosmic-ray electrons (CRe), subject to gains and losses using a Fokker-Planck solver. In particular, we included the effects of adiabatic stochastic acceleration (ASA) which is caused by the stochastic interaction of cosmic rays with diffusing magnetic field lines in super-Alfvenic turbulence. Moreover, we included shock acceleration and the seeding of CRe by galaxies.}  
{Our simulations generate cluster-scale radio sources during mergers, with properties that are in agreement with those observed for real radio halos. Furthermore, we find evidence of additional emission on larger scales. This emission resembles the radial distribution and the spectrum of a mega radio halo, but only when viewed close to the merger axis.}{In our simulation, the  mechanism responsible for the formation of diffuse radio emission, both in the form of classical and mega radio halos, is cosmic-ray re-acceleration by turbulence. This turbulence is more solenoidal and more subsonic in the classical radio halo region, than in the mega radio halo region.}


\maketitle

\section{Introduction  } \label{sec:introduction}

For at least three decades, it has been known that galaxy clusters host diffuse radio sources that are not associated with individual cluster galaxies \citep[e.g.][for a review]{2019SSRv..215...16V}. Such extended radio sources are associated with a perturbed dynamical state of the host cluster, and probe the existence of cosmic rays and magnetic fields in the intracluster medium (ICM) \citep[e.g.][for a review]{bj14}.
The unprecedented sensitivity to large-scale emission provided by the Low Frequency Array (LOFAR, \citealt{2013A&A...556A...2V}) has led to the discovery of very extended radio emission in clusters and between pairs of clusters \citep[e.g.][]{2019Sci...364..981G,2020MNRAS.499L..11B,2021A&A...648A..11O,2021A&A...654A..68H,bonafede21}. 
\citet{2022SciA....8.7623B} reported the LOFAR detection of diffuse radio emission in Abell 2255, covering almost the entire cluster volume, and containing classical radio halos and radio relics.
At the same time, \citet{2022Natur.609..911C} discovered a new class of diffuse radio sources beyond the volume of classical radio halos, dubbed mega radio halos (hereafter MegaRHs, in contrast to the ClassicalRHs introduced before). 
In four clusters, the ClassicalRH was found to be embedded in a much larger source that extends to scales of 2-3 Mpc, filling the volume of the cluster, at least up to $R_{500}$ \footnote{Here with $R_{500}$ and elsewhere in the paper with $R_{200}$ and $R_{100}$, we mean the radius enclosing a total matter overdensity $500$, $200$, and $100$ times larger than the critical cosmic one. Furthermore, $M_{500}$, $M_{200}$, and $M_{100}$ correspondingly denote the total masses measured within the same radii.}
The radial profiles of the radio emission in these newly discovered MegaRHs show two distinct components: a brighter core region that corresponds to the ClassicalRH and, separated by a clear discontinuity, a more extended and shallower low-surface brightness component. 
Based on the small number of known  MegaRHs, the transition from the ClassicalRHs to MegaRHs occurs at a radius that is about a half of $R_{\rm 500}$ from the cluster centre. The average emissivity of the MegaRHs is $\sim 20-25$ times lower than the emissivity of ClassicalRHs. For two MegaRHs, the combination of 50 and 140 MHz LOFAR observations yielded a relatively steep spectrum with a spectral index $\alpha\sim-1.6$.

MegaRHs show promise to reveal the physical properties of the outskirts of galaxy clusters, in a $\sim 30$ times larger volume than what has been possible to date. 
They may also help understand particle acceleration and magnetic field amplification in very weakly collisional plasmas 
\citep[e.g.][]{bl11,bj14,2016PNAS..113.3950R, PhysRevLett.131.055201,zhou2023magnetogenesis, 2018ApJ...863L..25S,2022hxga.book...56K}, as well as to better explore the connection between reacceleration and electron injection by quasi perpendicular shocks in the ICM \citep[ICM; e.g.][]{guo14a,ha21,boula24}.

Cluster-wide radio emission morphologies are compatible with a 
scenario where relativistic electrons are re-accelerated by second-order Fermi mechanisms in the turbulent ICM \citep[e.g.][]{gb01,2003ApJ...584..190F,cassano05,2016MNRAS.458.2584B}. In MegaRH, in particular, their steep radio spectrum makes this explanation by far the most plausible \citep[][]{2022Natur.609..911C,2023A&A...678L...8B,nishiwaki24}.

The theoretical challenge here is to maintain the relativistic electrons at energies that keep their synchrotron emission detectable at radio frequencies. The discontinuity in the surface brightness profile suggests a change either in the macrophysical (e.g. acceleration mechanisms) or in the microphysical properties of the plasma (e.g. acceleration efficiency, mean free path) when moving from the ClassicalRH to the MegaRH region. 
The scenario that we explore here is that both MegaRH and ClassicalRH are powered by turbulence \citep[e.g.][]{br07,bj14}, but their observable difference follows the significantly different properties of turbulence in the two regions.  
Previous studies, using cosmological simulations to interpret real observations \citep[][]{2022SciA....8.7623B,nishiwaki24}, suggest that radio emission in peripheral cluster regions require several percent of dissipation of kinetic turbulent energy for the amplification of magnetic fields and for the acceleration of relativistic electrons. This can happen if 
the turbulence in the MegaRH region has different properties than in the ClassicalRH region, and this work is also devoted to exploring these possible differences.\\

In a first attempt \citep[][hereafter Paper I]{2023A&A...678L...8B}, we 
investigated the lifecycle of relativistic electrons in the outskirts of galaxy clusters under realistic conditions. We calculated under which circumstances the observed volume of MegaRHs could be filled with radio-emitting electrons. 
However, in this first work we did not model the full ageing and re-acceleration process of CRe mixed with the ICM, nor did we simulate their synchrotron emission in any detail. Here we extend our previous work in order to quantitatively test the theoretical formation scenarios of MegaRH.

A few other papers recently explored the formation of very extended radio emission in simulated clusters of galaxies, albeit with a more simplified approach than the one we have used for this work.
By relying on the classical transit time damping (TTD) acceleration (in which the re-acceleration is powered by magneto-sonic waves excited by the compressive part of the turbulent kinetic power; e.g. \citealt[][]{cassano05,br07,2013MNRAS.429.3564D}), \citet{boss23} simulated the emergence of diffuse radio emission in a suite of constrained simulations of the Local Universe to study the production of radio emission from turbulent re-acceleration in massive clusters, such as Coma.  
On the other hand,
\citet{nishiwaki24} explored the adiabatic-stochastic-(re-)acceleration scenario  \citep[][]{bl11,bv20} and 
used a Fokker-Planck method to evolve the distribution of relativistic electrons in a perturbed cluster from the same suite of magnetohydrodynamics (MHD) cosmological simulations used in our Paper I. 
Limited to the analysis of a single cluster snapshot from the simulation, they found  that if a conservative fraction ($\sim 1 \%$) of solenoidal turbulent energy is dissipated for magnetic energy and electron re-acceleration, a large fraction of peripheral cluster volume can produce detectable radio emission. 

In this work, we apply a full Fokker-Planck solver over a population of Lagrangian tracers. This will be done for about a hundred root-grid time steps (approximately yielding a $\Delta t=100 ~\rm Myr$ time step) of the simulation, in order to test whether the complex cycles of (re-)acceleration experienced by relativistic electrons can reproduce observational properties of MegaRH.\\

Our paper is structured as follows:
In Section 2 we introduce our cosmological simulations and the post-processing numerical methods to evolve relativistic electrons. In Section 3  we present our results, including the observable properties of the simulated radio emission. We also discuss the physical and numerical uncertainties of our model. Finally, in Section 4, we present our conclusions. 

\section{Methods and simulations}  \label{sec:methods_sim}

\subsection{Cosmological simulation: ENZO} \label{subsec:cluster_evol}

Continuing on from Paper I, we analysed the same cosmological simulation of a cluster of galaxies, produced with {\enzo} code \citep[][]{enzo14}, for which our previous work predicted significant chances of re-acceleration of relativistic electrons, in both the Classical and Mega RH regions. The simulation includes the self-gravity for ordinary and dark matter, radiative equilibrium cooling, but no other galaxy formation-related physics. 

No feedback processes were needed as the resolution is not high enough to produce strong cooling flow, and the cluster is undergoing frequent merging to keep it from cooling castrophically.
Magnetic fields are evolved using ideal MHD using the hyperbolic cleaning method \citep[][]{enzo14}. At the beginning of the simulation, at a redshift of $z=40$, we injected a uniform magnetic field of $\rm B_0=0.4$ nG in all directions. The most massive cluster at $z=0$ has a mass of $\rm M_{100} = 3.8\times 10^{14}\ M_{\odot}$ and a virial radius of $\rm R_{100} = 1.5$ Mpc. The simulation has been performed on a computational grid of $128^3$ cells, namely $(100\rm~ Mpc/h^{-1})^3$, with dark matter particles ($\rm m_{DM0}=3\times 10^{10} M_{\odot}$ of mass resolution).
A zoomed-in region where the cluster forms (with size $(16\, {\rm Mpc})^3$) was further refined with four additional levels of $\times 2$ refinement in spatial resolution each (and $\times 2^3$ refinement in mass resolution of the dark matter component, at each refinement step) with nested regions of decreasing size. The entire cluster region is thus resolved with a maximum dark matter mass resolution of $m_{\rm DM}=7.3 \times 10^6 M_{\odot}$ (i.e. $m_{\rm DM0}/8^4$), and with a uniform spatial resolution of $70$ kpc/cell. On top of this, we also allowed for two extra levels of adaptive mesh refinement within the innermost region by refining on local gas or dark matter overdensities, up to a final maximum resolution of $\approx 12 \rm ~kpc/h= 18$~kpc (comoving). 

This particular cluster is the same as used in our Paper I, only that we now use an upgraded Fokker-Planck sover for the more advanced treatment of the cosmic-ray evolution. This cluster is smaller than the clusters for which MegaRH have been detected (i.e. $3\sim 10^{14} \rm M_{\odot}$ vs $\sim 10^{15} \rm M_{\odot}$, yet the amount of simulated snapshots needed to model the evolution of CRe momenta in post-processing with our approach is already large enough ($\sim 200 \rm ~Tb$) to pose a significant challenge.
The application to more massive clusters will be explored in a following campaign of dedicated simulations. 

The assumed cosmology is a flat $\Lambda$CDM model with $H_0 = 67.8\, \rm  km/s/Mpc$, $\Omega_M=0.308$,  $\Omega_\Lambda=0.692$, $\Omega_b=0.0478$, and $\sigma_8=0.811$. 
 
\begin{figure*}[h!]
    \centering
     \includegraphics[width=0.99\textwidth]{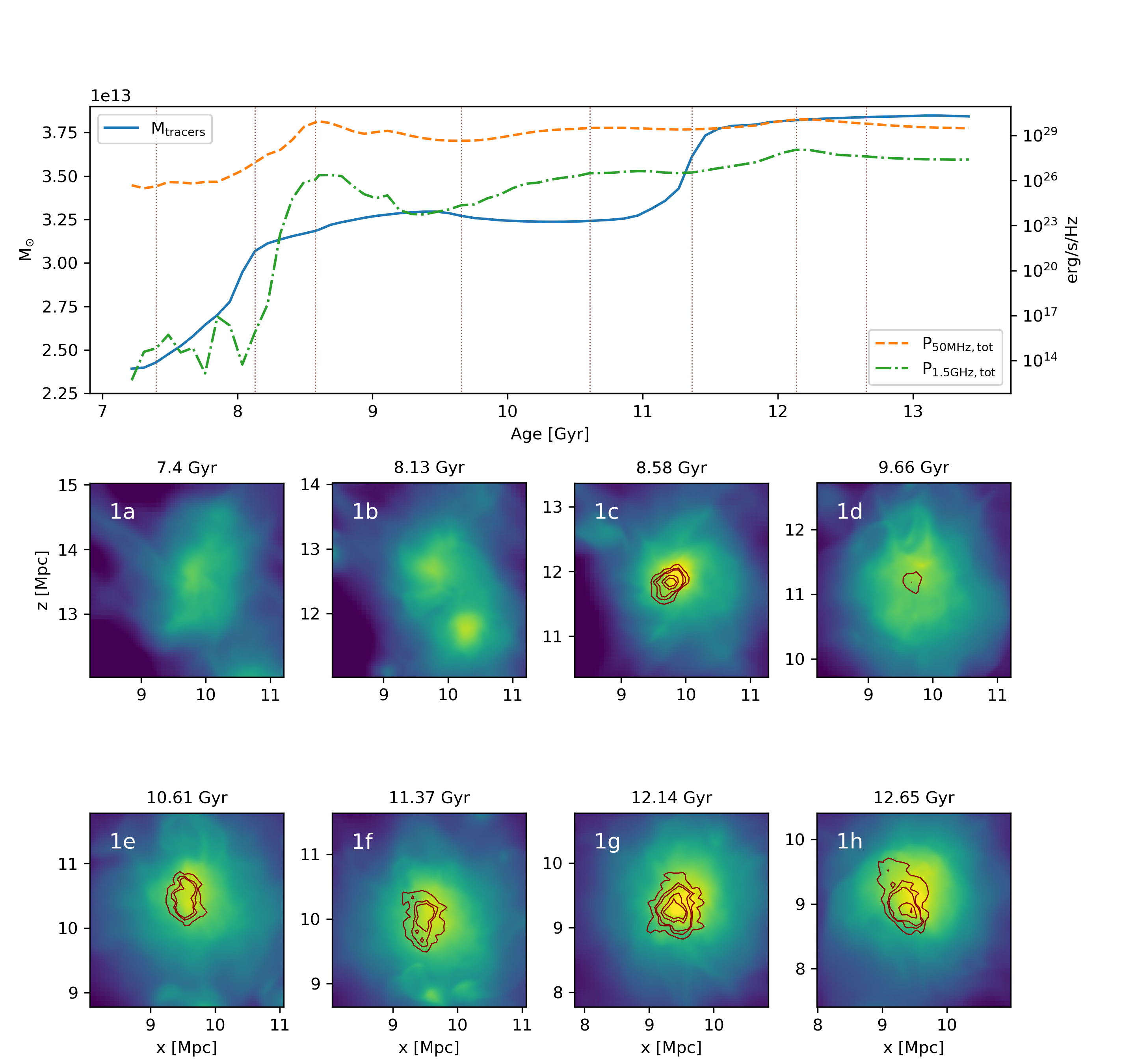}
    \caption{Representative view of the cluster evolution during the simulation. The upper panel shows the total mass of the tracers within 1.5 Mpc from the centre of the cluster, as a function of time. Overplotted is the synchrotron emission within the simulation at $\rm \nu = 50\ MHz$ and $\rm \nu = 1.5\ GHz$. The lower panels show the projected gas density maps, with contours encompassing the emission from $\rm P_{150MHz}\ \geq 10^{26}$ erg/s/Hz/pixel, spaced by $10^{0.5}$ erg/s/Hz/pixel, where the size of each pixel is 39", assuming the cluster is placed at $z=0.095$. Vertical dotted lines drawn in the upper panel indicate the ages corresponding to the time depicted in the maps.}
    \label{fig:mass}
\end{figure*}

\begin{figure*} 
    \includegraphics[width=0.33\textwidth]{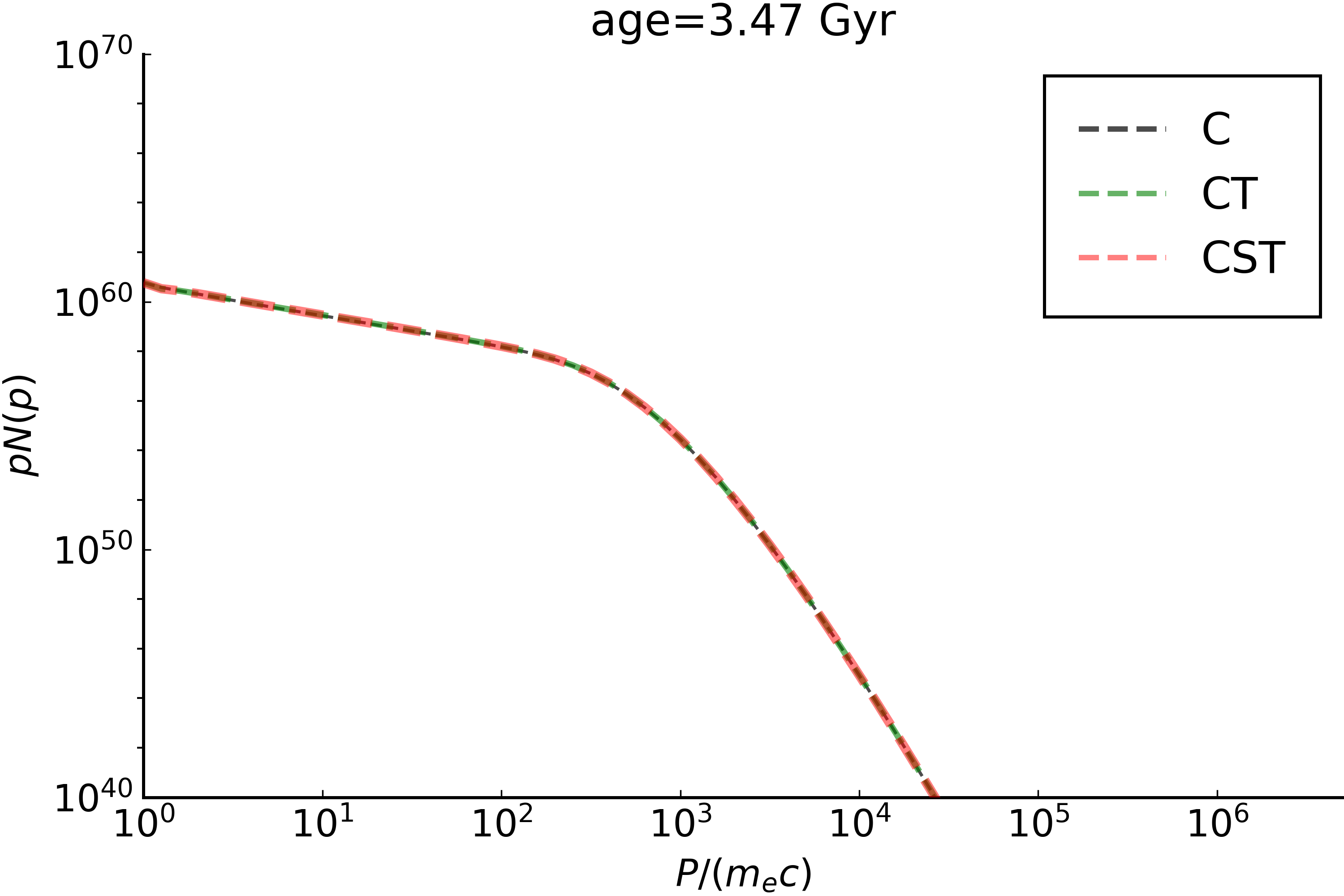}
    \includegraphics[width=0.33\textwidth]{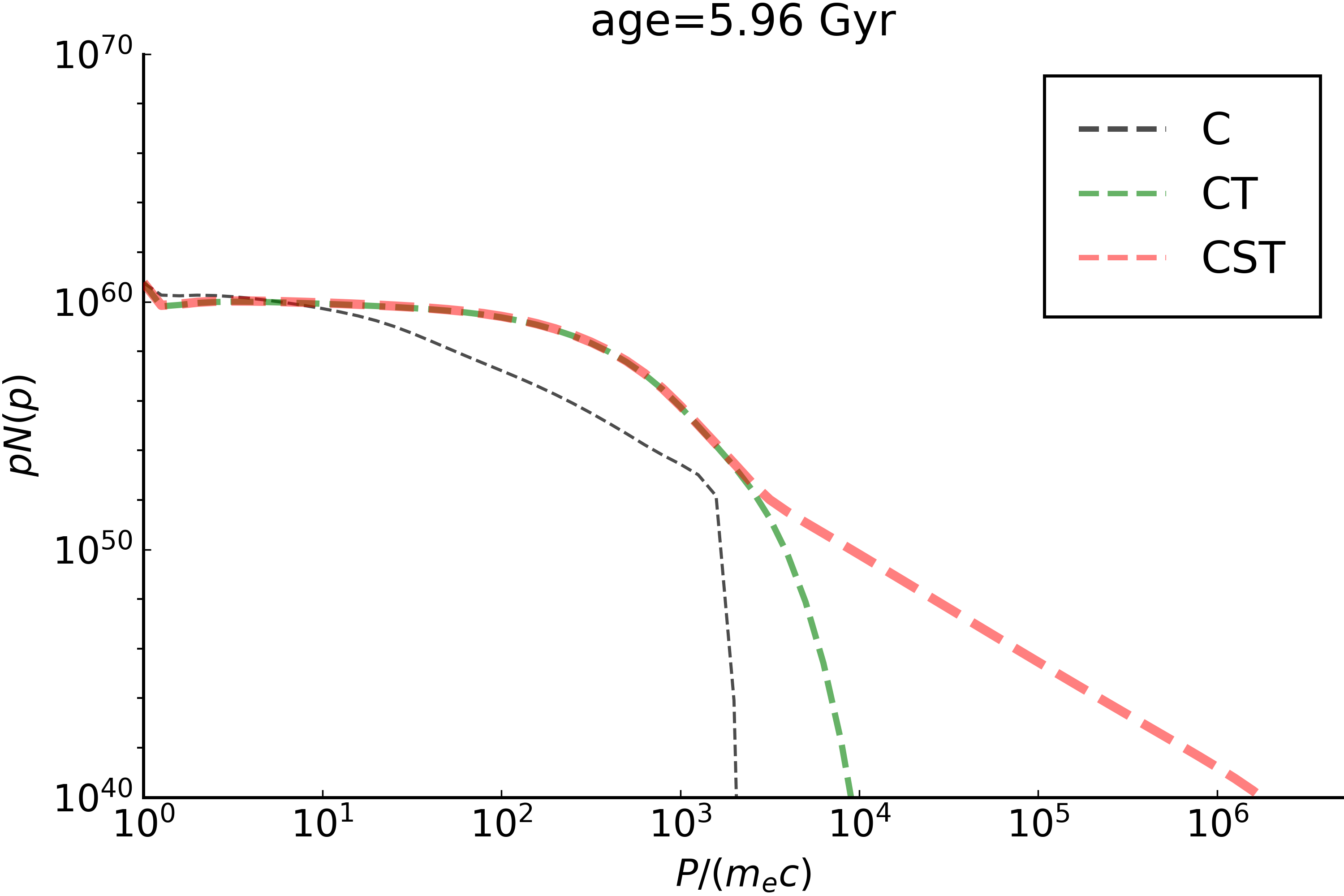}
    \includegraphics[width=0.33\textwidth]{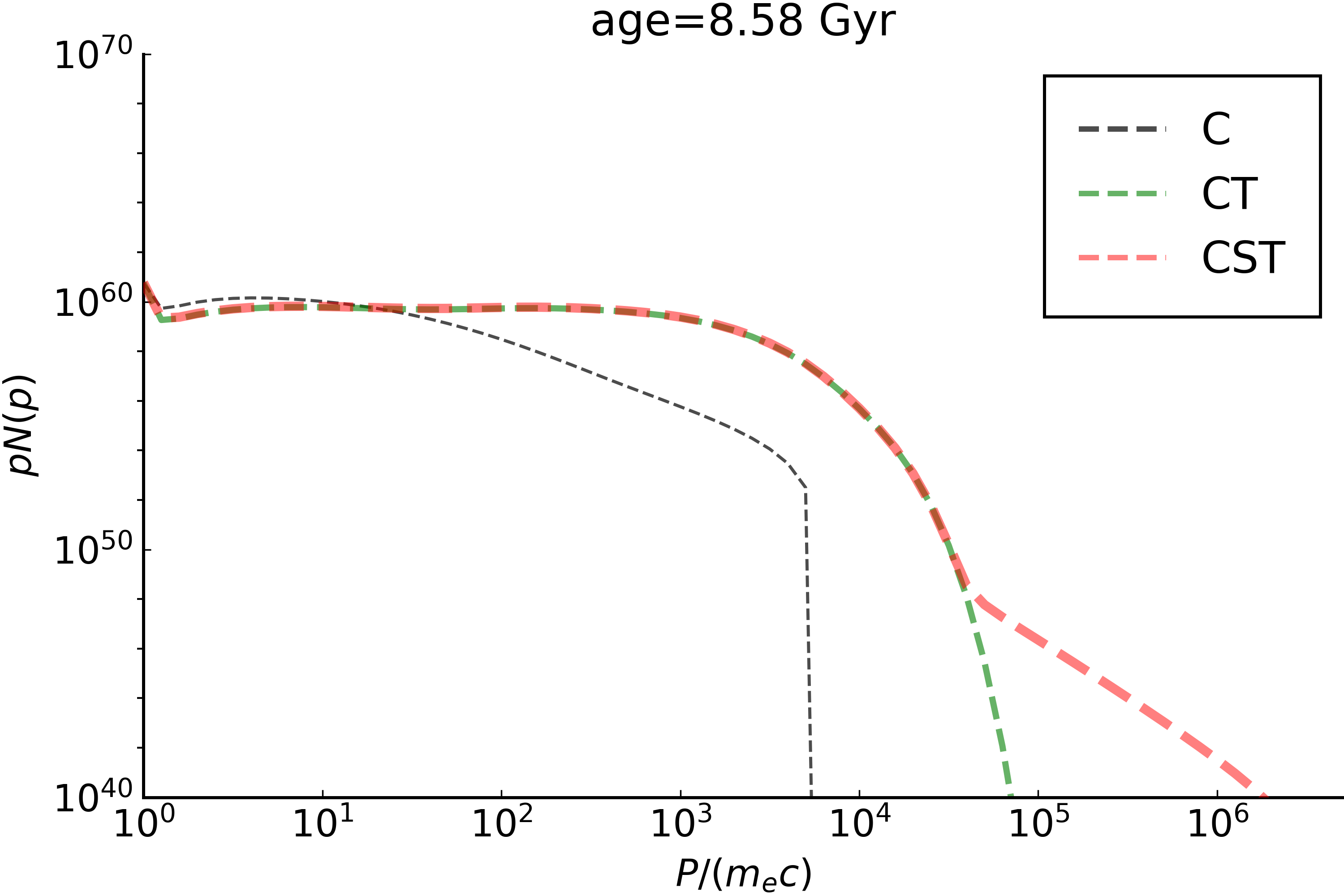}
    \includegraphics[width=0.33\textwidth]{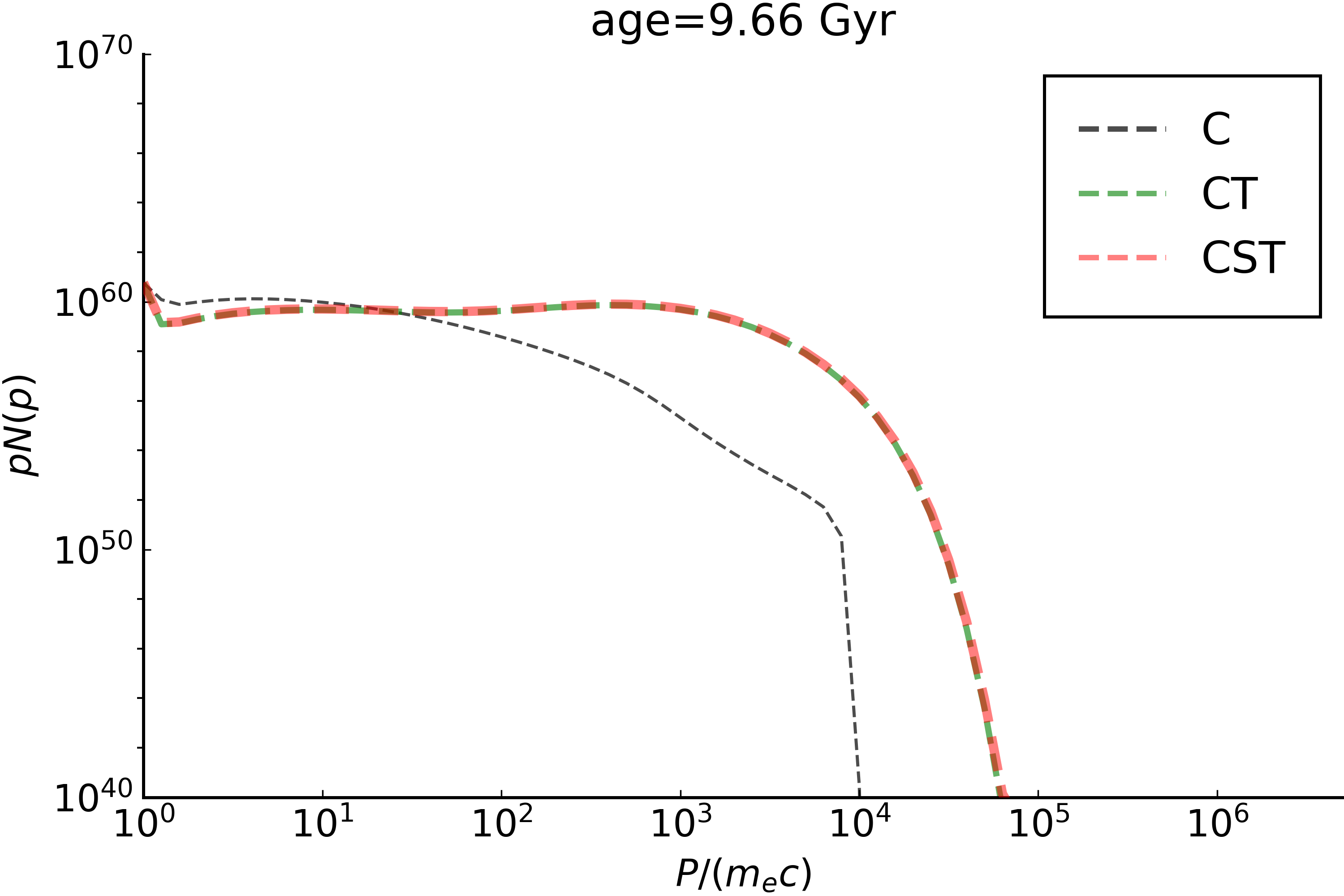}
    \includegraphics[width=0.33\textwidth]{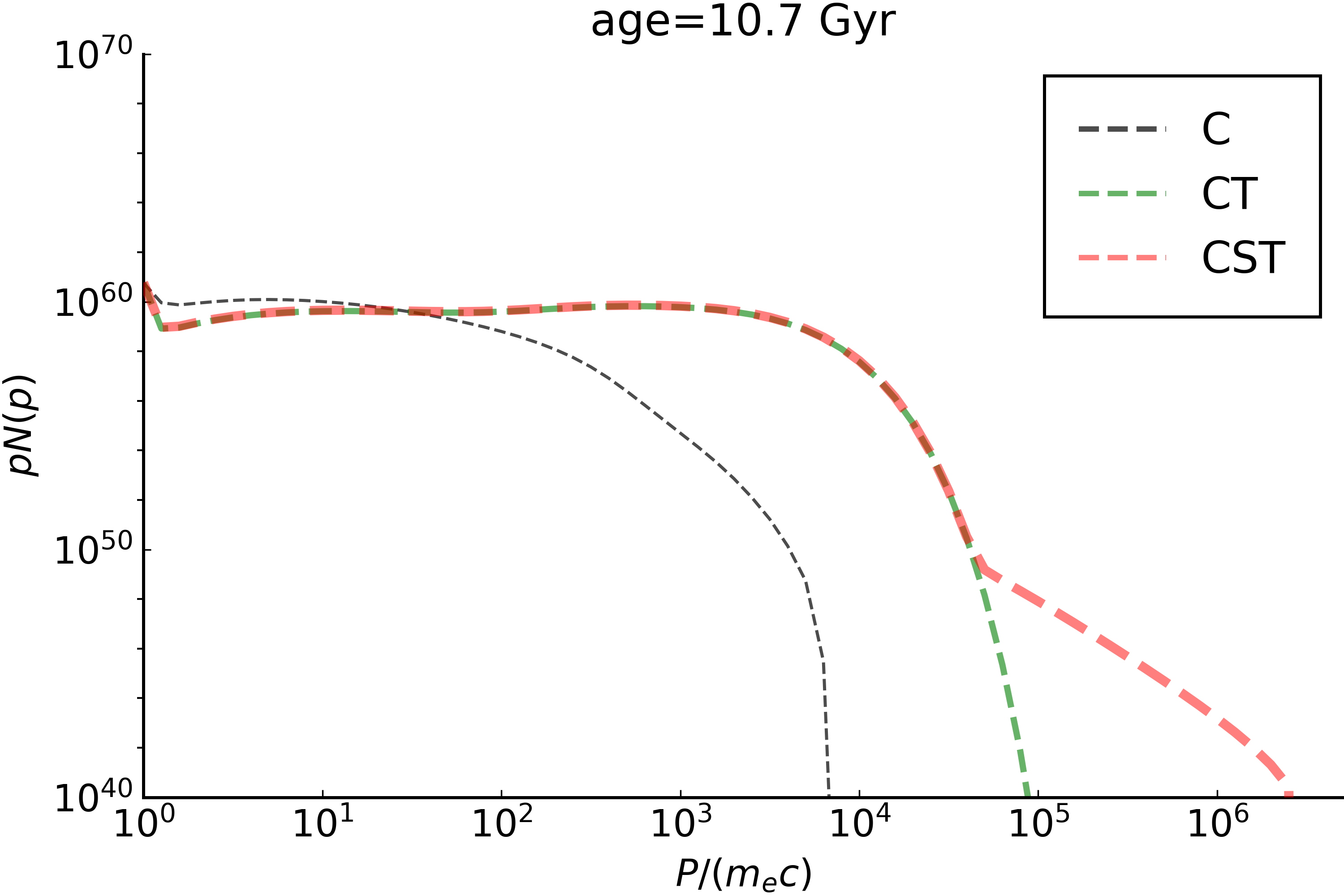}
    \includegraphics[width=0.33\textwidth]{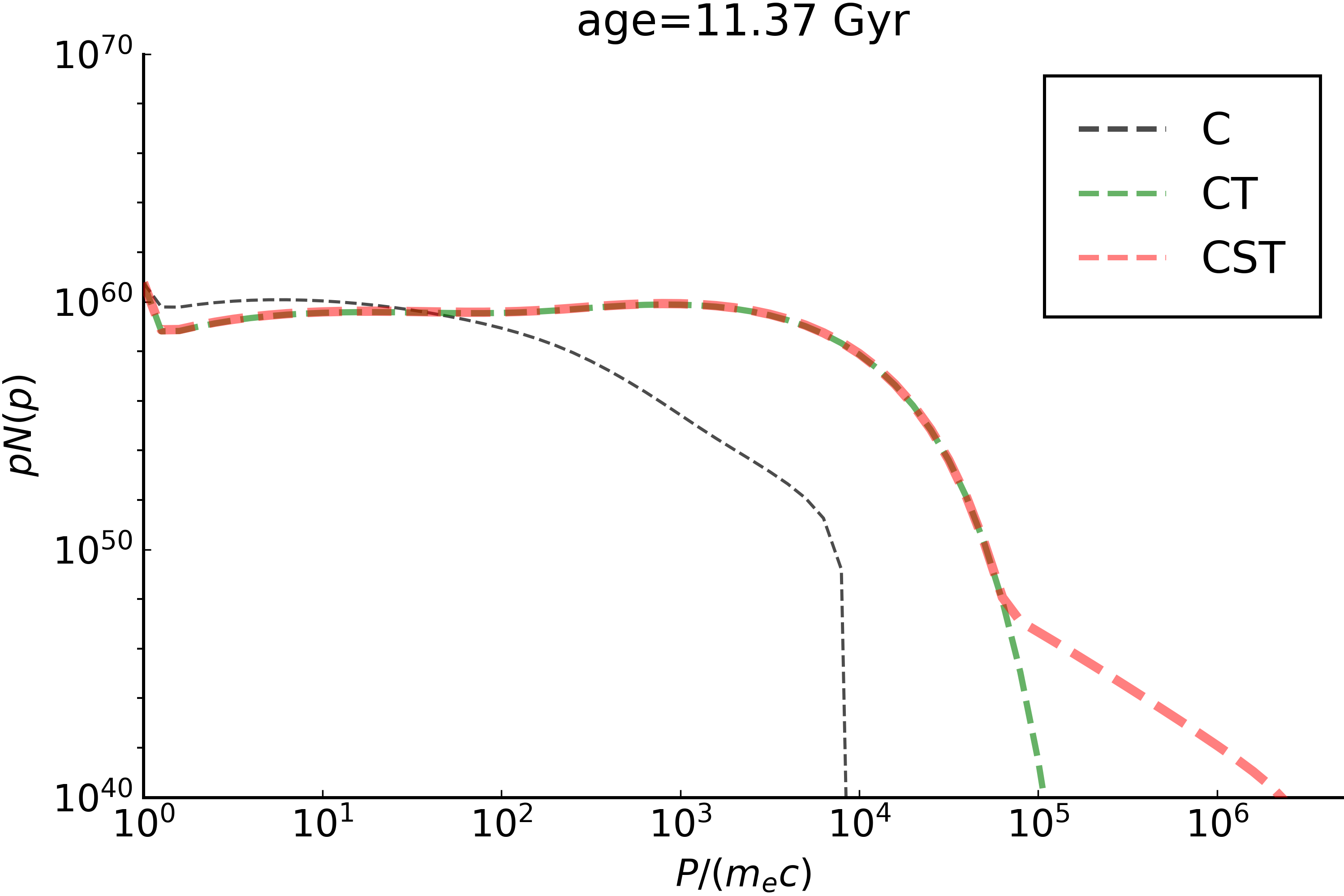}
    \includegraphics[width=0.33\textwidth]{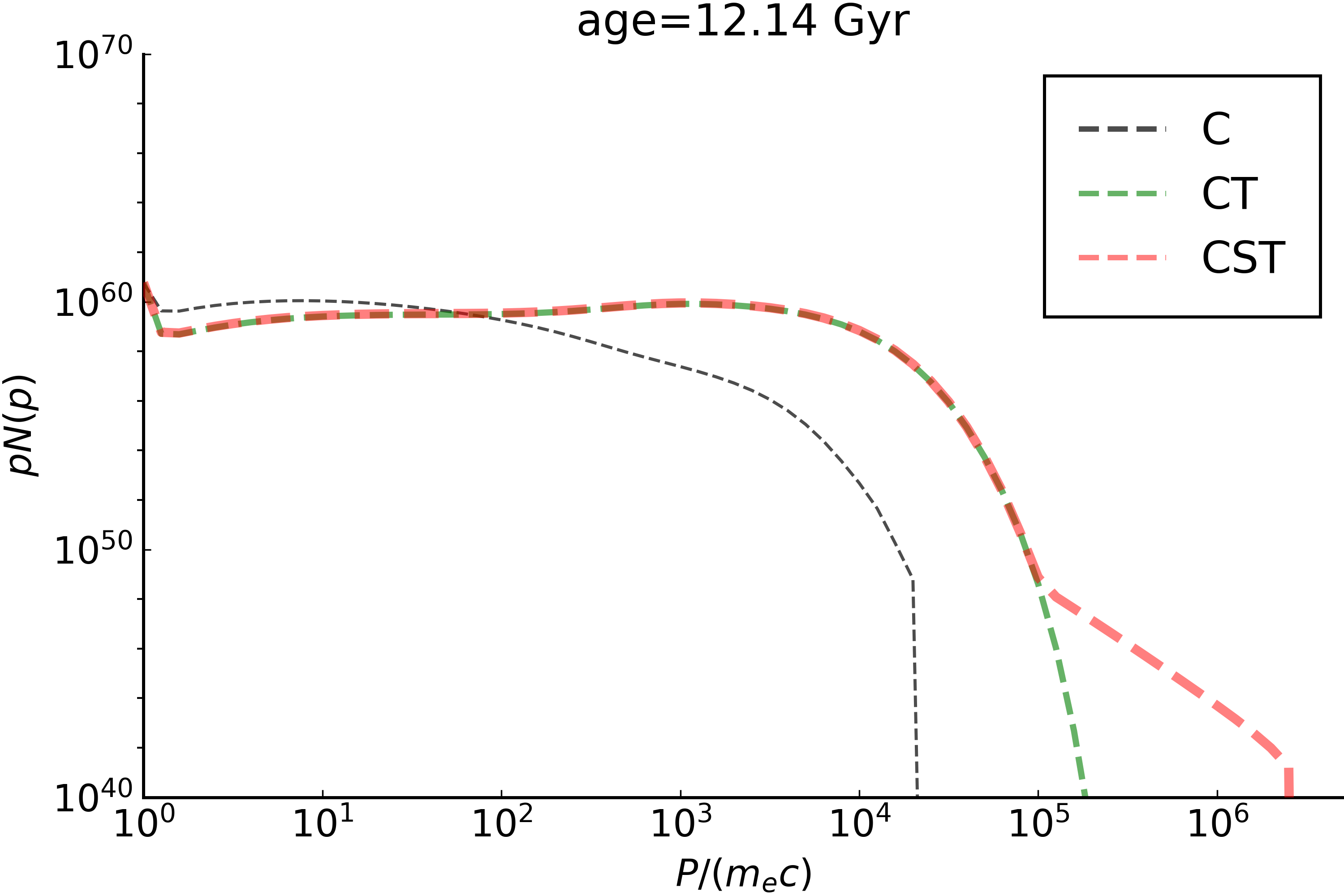}
    \includegraphics[width=0.33\textwidth]{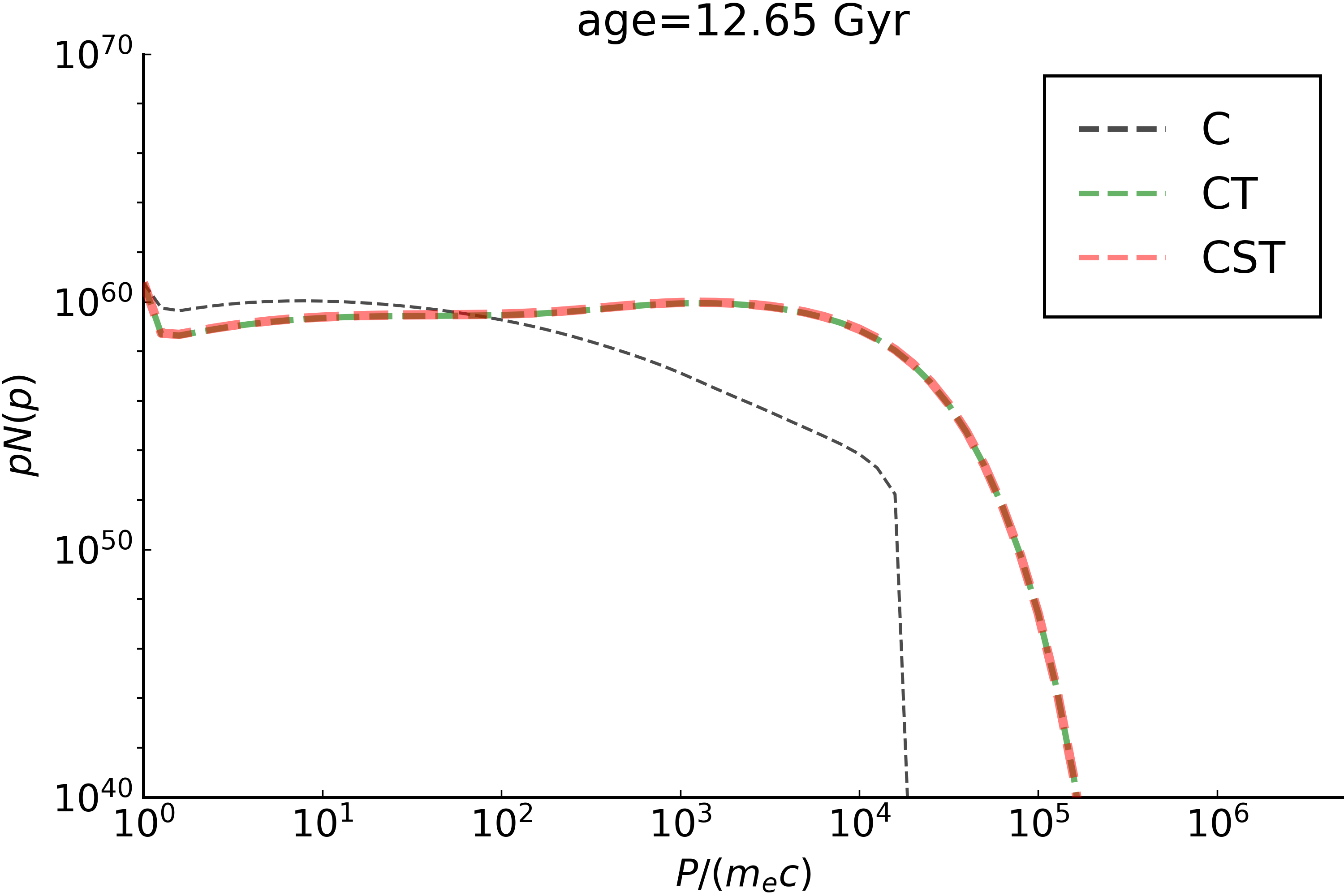}
    \caption{Evolving momentum spectra of all populations of relativistic electrons, for three physical models (C=only cooling; CT=cooling and turbulent re-acceleration; CST=cooling, turbulent-reacceleration and shock acceleration).  
    }
    \label{fig:roger_spt}
\end{figure*}

\subsection{Simulations of cosmic-ray electron spectra: CRATER \& ROGER}\label{subsec: crater_roger}

We modelled the spatial evolution of relativistic electrons in post-processing by sampling of a distribution of discrete Lagrangian particles, created  by sampling a fixed gas mass resolution (approximately of $5 \cdot 10^8 M_{\odot}$) in the grid of density values produced by the ENZO simulation. Once injected, the trajectories of tracers are evolved forward in time, with a simple time integrator $\vec{r}(t+dt) = \vec{r}(t) + \vec{v_{\rm gas}} dt $ (where $\vec{r}(t)$ is the 3-dimensional position as a function of time and $\vec{v_{\rm gas}}$ is the Eulerian gas velocity referred to the position of the tracer), after interpolating the 3d-velocity field with a  cloud-in-cell spatial interpolation method, as explained in \citet{wi17}.
To each Lagrangian particle, we attach energy spectra discretised in logarithmic momentum bins, which we evolved with a Fokker-Planck approach to solve:

\begin{align} 
\label{fp}
    \frac{\partial N(p)}{\partial t} &= \frac{\partial}{\partial p} \left[ H(p) N(p)  + D_{\mathrm{pp}} \frac{\partial N(p)}{\partial p} \right] + Q_{\mathrm{inj}}(p,t) ,
\end{align}\\
where $N(p)$ is the momentum distribution of electrons,
$Q_{\mathrm{inj}}(p,t)$ is a time-dependent source term for the injection of electrons (in our case, only allowed for diffusive shock acceleration; DSA) and 
$H(p)$ is the generalised cooling function $H(p)$: 

\begin{align}
    H(p) &= -\frac{2}{p}D_{\mathrm{pp}} + \sum_i \frac{dp}{dt} _{i} ,
\end{align}
which sums up all relevant losses for the particle population (as in Paper I: ionisation and Coulomb losses, adiabatic losses, synchrotron and Inverse Compton losses, e.g. \citealt{sa99}), as well as the gain by turbulent acceleration, via the $D_{\mathrm{pp}}$ term.
While we give a detailed explanation of how the $D_{\mathrm{pp}}$ turbulent re-acceleration term is computed below (as it is central to our study), we give an overview of the rest of the adopted cooling terms in Appendix \ref{appendix1}.
For injecting and advecting Lagrangian tracers, we used the CRaTer code by \citet{wi16}, which allows us to generate the trajectories of $10^5$ tracers. These are injected at redshift $z=4$ in the highest resolution region of the simulation, with each tracer sampling a gas mass resolution of  $5 \times 10^8 M_{\odot}$. The trajectory of tracers were updated using 110 snapshots recorded from the simulation, from $z=2$ to $z=0$.  For each tracer we recorded the gas density, temperature, magnetic field, divergence and the curl of velocity as well as the Mach number of shocks, using cloud-in-cell interpolation. The Mach number is calculated from the temperature jump recorded by tracers, as in our previous work.
To each tracer we assigned a simple initial power-law distribution of electrons in momentum space, that is, $N(p,t=0)=N_0 p^{-\delta}$, with $\delta=2$. The normalisation, $N_0$, was chosen in order for the total number of injected electrons to be $\phi=10^{-5}-10^{-7}$ of the density of thermal electrons. This range of values is compatible with the number density of cosmic ray electrons which can be inferred based on the extent and power of observed radio halos \citep[e.g.][for a recent review]{2024arXiv240316068V}.
We solved the subsequent time-dependent diffusion-loss equation of relativistic electrons with the parallel ROGER\footnote{\url{https://github.com/FrancoVazza/JULIA/tree/master/ROGER}} Fokker-Planck solver, which is implemented in the Julia programming language\footnote{\url{https://julialang.org}}. We used $N_b=100$ momentum bins equally spaced in $\rm log (p)$, in the $p_{\rm min} \leq p \leq p_{\rm max}$ momentum range (with $P=\gamma m_e v $ and $p=P/(m_e c)$ is the normalised momentum of electrons). In all production runs we used $p_{\rm min}=10$ and $p_{\rm max}=10^6$ (hence $d\log (p)=0.05$).
The details of the code are discussed in \citet{va23a}.\\

Compared to our previous work, we use an updated numerical solver to follow the stochastic, as well as the systematic, acceleration by Fermi-II turbulent acceleration \citet{1970JCoPh...6....1C}, i.e. the fact that the Fermi-II acceleration process can cause a rigid shift of the electron distribution in momentum space, as well as its diffusion in momentum space, which is important for producing radio tails. 
We followed the formalism by \citet{2013MNRAS.429.3564D}, with the important difference that we computed the $D_{\rm pp}$ coefficient based on the adiabatic-stochastic-acceleration model by \citet{2016MNRAS.458.2584B}.  We evolved the population of relativistic electrons considering radiative losses (synchrotron and Inverse Compton), Coulomb and ionisation losses, and adiabatic terms (e.g. compression and rarefaction). In our baseline model, the initial ratio between the number of relativistic electrons and of thermal electrons is $\phi=10^{-6}$. 

We also included the (re-)acceleration from DSA, based on the Mach numbers recorded by tracers which enters in the solved Fokker-Planck equations of Eq.~\ref{fp} as an additional source term which is only introduced whenever strong enough shocks are recorded by the tracers, $Q_{\rm inj}(p,t)$. 

The injection efficiency of CRe at shocks, especially in the low-Mach number regime is highly uncertain and cannot be derived from first principles \citep[e.g.][]{Bykov19}.
Moreover, more complex and realistic theoretical recipes to model the injection of electrons arising from the contribution of shock drift acceleration in low-Mach number and oblique shocks can be considered \citep[e.g.][]{2022ApJ...927..132A}, and thus our prescription for $Q_{\rm inj}(p,t)$ is meant to be a first, simplistic choice  to test our algorithm. Here we adopted the convenient parametrisation, given by \citet{2023MNRAS.519..548B}, to reproduce the DSA acceleration efficiency proposed by \citet{2019ApJ...883...60R}, rescaled for the CR electron-to-proton ratio ($K_{\rm ep}$),  which we fix by requiring an equal number density of supra-thermal CRe and protons above the injection momentum,  $K_{\rm ep}=[(m_p/m_e)^{(1-\delta_{\rm inj})/2}]^{-1}$, as in ~\cite{2013MNRAS.435.1061P}. Any additional dependence between shock acceleration efficiency and the angle between the shock normal and the upstream magnetic field is neglected for simplicity. Our previous work has shown that in the ICM the majority of merger shocks are quasi-perpendicular and suitable for maximal electron acceleration ~\citep[e.g.][]{wittor20,Banfi20}. 
While this Fokker-Planck method, with the new implementation of the stochastic Fermi II acceleration based on \citet{2013MNRAS.429.3564D} written and parallelised in Julia language, will be presented in more detailed forthcoming paper (Vazza et al, in prep.), we notice that the basic code has been already documented in our previous publications on this line of research \citep[][]{va21jets,va23a,va23b,2023MNRAS.526.4234S}.

We simulate the effect of Fermi-II re-acceleration via ASA model \citep[][]{2016MNRAS.458.2584B}. This mechanism stems from the stochastic interaction of cosmic rays with diffusing magnetic field lines in super-Alfvenic turbulence. The acceleration rate used in the Fokker-Planck equation, $dp/dt$, depends on the solenoidal turbulent energy flux, $F_{\rm turb} = \frac{\rho \delta V^3_{\rm turb}}{2L}$, which is conserved in the Kolmogorov model of turbulence. As in Paper I, we use the gas vorticity to estimate the local level of solenoidal turbulence, $\delta V_{\rm turb}= |\nabla \times \vec{v}| L$, where $L\approx 54 \rm ~kpc$. Now we write, $dp/dt \approx 4~D_{\rm pp}/p$,  where $D_{\rm pp} \approx (54 F_{\rm turb} ~p^2)/(c~\rho ~V_A)$, $\rho$ is the local gas density and $V_A$ is the local Alfven wave velocity, following \cite{bv20}. 

We compensate for the effect of the finite numerical resolution of our simulation, which limits the small-scale dynamo amplification of magnetic fields, with a physically motivated rescaling of the magnetic fields strength associated with each tracer. In particular, we assumed that a fixed fraction $\eta_B$ ($=2-5\%$, see tests below) of the energy flux of turbulent kinetic energy gets dissipated for the amplification of magnetic fields, as in \citet[][]{bm16}. To each tracer particle we assign the maximum value between the magnetic field given by the direct MHD simulation, and $B_{\rm turb}^2/8\pi \sim \eta_B F_{\rm turb} \tau \sim {1 \over 2} \eta_B \rho \delta V_{\rm turb}^2$, where $\tau= L/\delta V_{\rm turb}$ is the turnover time. This magnetic field amplitude is used to compute synchrotron losses and the Fermi-II re-acceleration rate. 

Our Fokker-Planck method evolves the spectra of all tracers with a much more refined time stepping than for the advection of tracers. Typically we employ $\sim 100$ sub-cycling time steps between two time steps of the cosmological simulation, the exact number being set by the minimum between the acceleration and the loss timescales for particles with $\gamma \sim 10^4$, which contribute the most to the radio emission.
Moreover, we used a limiter for the maximum turbulent velocity used for the Fermi-II acceleration, on the basis that in the presence of occasional complex gas velocity patterns, induced by oblique shocks and turbulence, the vorticity can become a biased proxy for solenoidal turbulence. In order not to keep a conservative value of the $D_{pp}$ re-acceleration coefficient, we set $\delta V_{\rm turb}/c_s \leq 0.4$ as a a maximum amplitude of the root mean square (rms) turbulent velocity used in the Fokker-Planck method. 

Finally, to compute the synchrotron emission for all tracers at 50, 120, 610, 1400 and 5000 MHz, we employed fitting functions for the synchrotron emissivity \citet{2014MNRAS.442..979F}. 

The ROGER simulations allow us to evolve in parallel the following three different models for the evolution of relativistic electrons:

\begin{itemize}
\item C: a scenario in which electrons, after their injection, only evolve via adiabatic, collisional and radiative losses, and compression gains; 
\item CT: a scenario in which also turbulent re-acceleration is acting on the electron spectra;
\item CST: a scenario in which, both, turbulent re-acceleration and shock re-acceleration are active, as well as the injection of new electrons through DSA.
\end{itemize}

\begin{figure*}
    \centering
    \sidecaption
    \includegraphics[width=12cm]{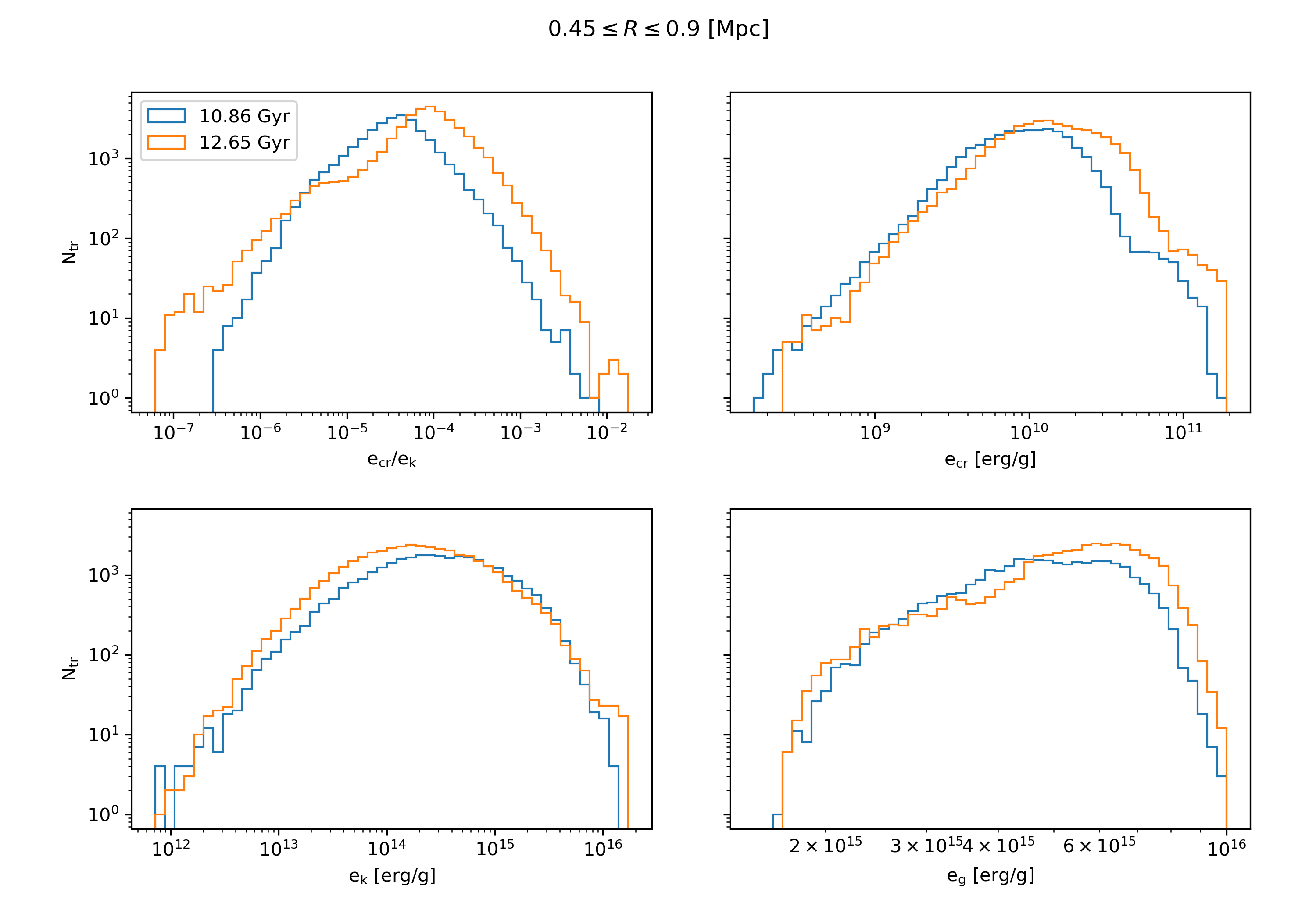}
    \caption{Energy distributions of the tracers inside $R_{500}$ at two different redshifts, marking an epoch before ($10.86$ Gyr) and after ($12.65$ Gyr) the last important merger in the system, and the onset of diffuse radio emission. Top-left: distribution of the ratio between the CR energy and the turbulent energy. The ratio increases with time and broadens after the merger; top-right: distribution of the CRe energy; bottom-left:  kinetic, or turbulent, energy distribution; bottom-right: the gas energy distribution.}  
    \label{fig:energy_distribution}
\end{figure*}

\section{Results} \label{sec:results}

\subsection{Cluster evolution} 

Fig.~\ref{fig:mass} shows the evolution of the total gas mass, measured using tracers, within the $R_{\rm 100}$ of the cluster. Overlaid is the radio emission at two different frequencies. The lower panels show the evolution of the projected gas density, with the radio emission contours overlaid.

In the first phase of its evolution, up to $7 \rm ~Gyr$ (panel 1a in the figure) since the start of the simulation, the cluster has accreted a large population of smaller clumps. Its shape is still asymmetric and perturbed. The merger between $\sim 8$  and $\sim 9 \rm ~Gyr$ (panels 1b to 1d) finally produces a massive cluster in this region, with a mass that is approximately $\sim 85\%$ of the total final mass of the cluster at $z=0.0$.

From this point on, the cluster will still undergo several minor mergers, and a larger one, approximately with a mass ratio between 1:5 to 1:6 (we analyse the role of the merger mass ratio in Sec.\ref{sec:conclusions}). The latter is the most powerful event in terms of the injection of turbulence and for the production of diffuse radio emission. 
First a small satellite (which was entered the cluster about $1 \rm ~Gyr$ before) hits the cluster core at about $10.6 \rm ~Gyr$, and it creates asymmetries in the gas distribution, and it launches merger shock waves through the ICM.  
Again, between $\sim 11$ and $\sim 12 \rm ~Gyr$, the last major merger in the cluster launches several shocks (panel 1f). Soon after that, this merger distributes a large fraction of the gas from the inner part of the cluster to larger radii.  At the same time this event generates significant turbulence. The formation of radio emission in our model, including the MegaRH-like structure out to $\rm R_{500}$, is causally connected to this event, but the radio emission peaks a few hundreds of million years later than this, owing to the fact that turbulent energy cascade only after an eddy turnover time. Moreover, the most spectacular effect here is in the production of extended diffuse emission beyond the innermost $\sim \rm Mpc^3$ occupied by the Classical RH emission, which we shall discuss in the next Sections.


\begin{figure*}
    \centering
    \includegraphics[width=0.95\textwidth]{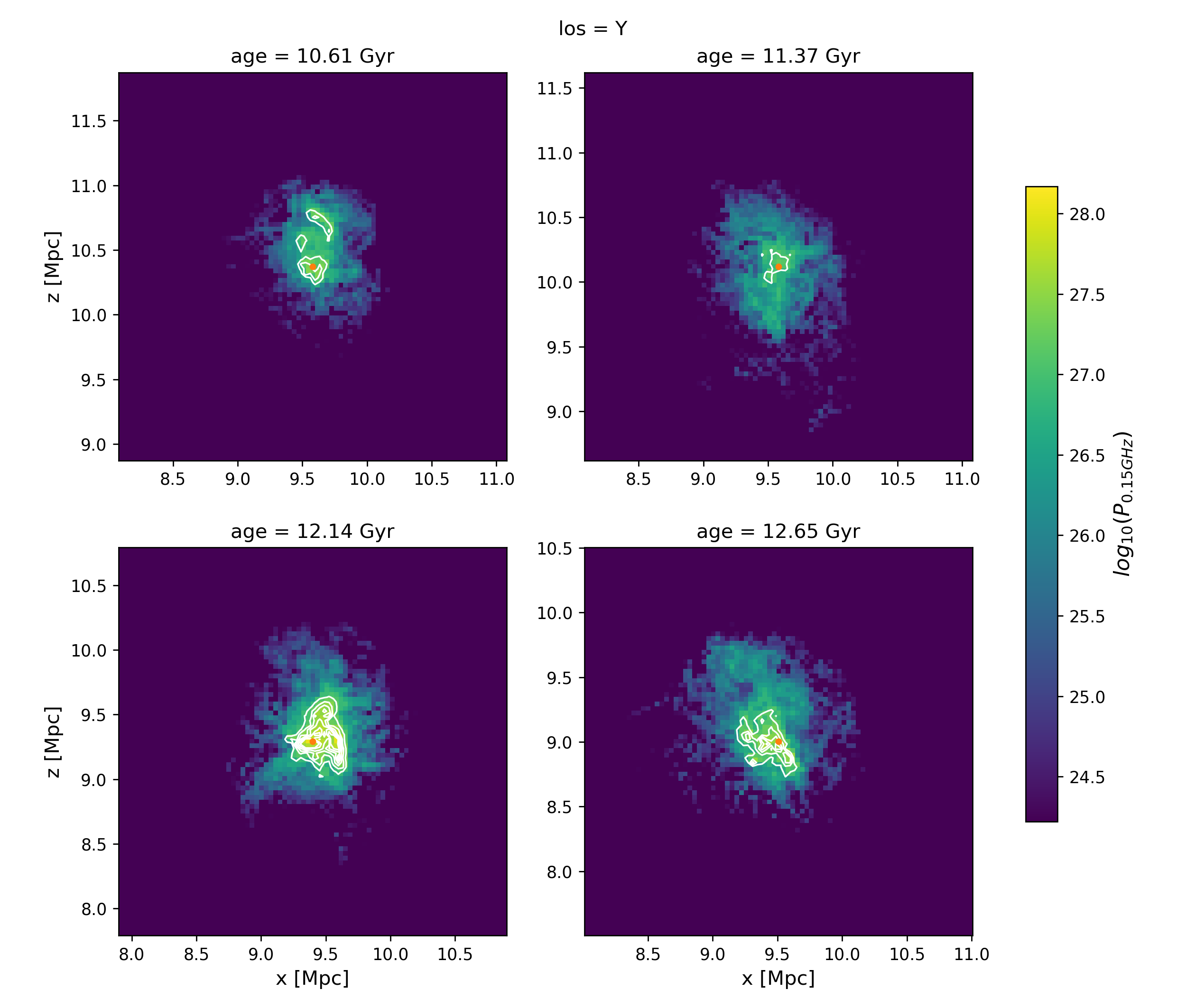}
    \caption{Synchrotron radio emission maps at 150MHz for four epochs already considered in previous plots, in units of [$\rm erg/s/Hz/pixel$], where the size of each pixel is 39". 
    The peak in each map is marked by the orange dot and is the centre used to compute the radial emission profiles, while the white contours represent the noise levels at the different levels of $\rm \sigma_{\rm rms}$.}
    \label{fig:4snaps}
\end{figure*}

\begin{figure*}
    \centering
    \includegraphics[width=0.8
\textwidth]{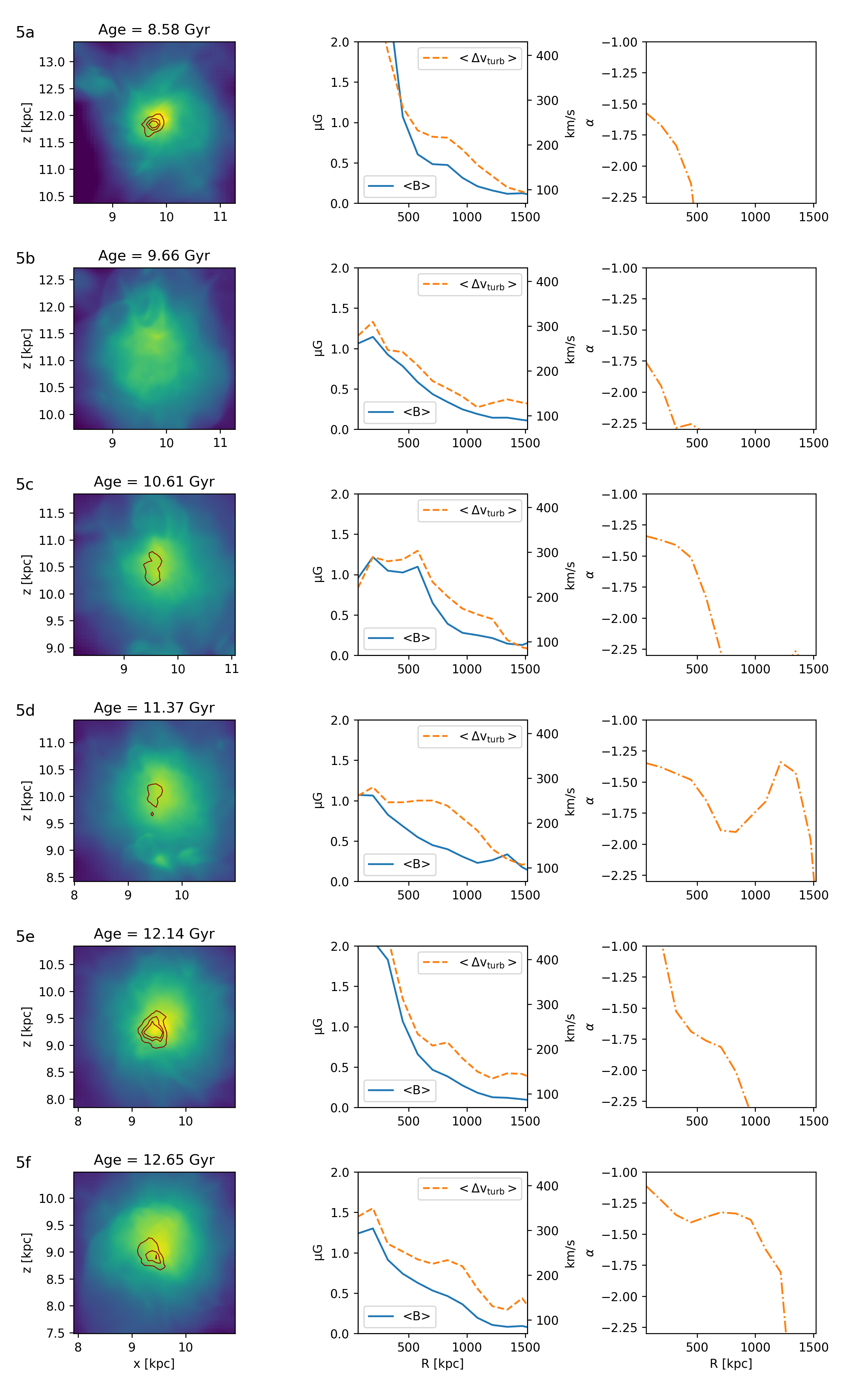}
    \caption{Each row shows the map of density and synchrotron emission, the magnetic field strength, and the turbulent velocity, both averaged over an annulus of 128 kpc, and the radially averaged spectral index. All the quantities are taken over the projection along the $y$-axis. The contours represent the radio power at $\rm P_{\rm 150MHz}\ =\ \{10^{27},\ 5\cdot10^{27},\ 10^{28}\} \ [erg/s/Hz/pixel]$, where the size of each pixel is 39", assuming the cluster is always observed at $z_{\rm obs}=0.095$.}
    \label{fig:6_maps_b_turb_alpha}
\end{figure*}

\begin{figure*}
    \centering
    \sidecaption
    \includegraphics[width=12cm]{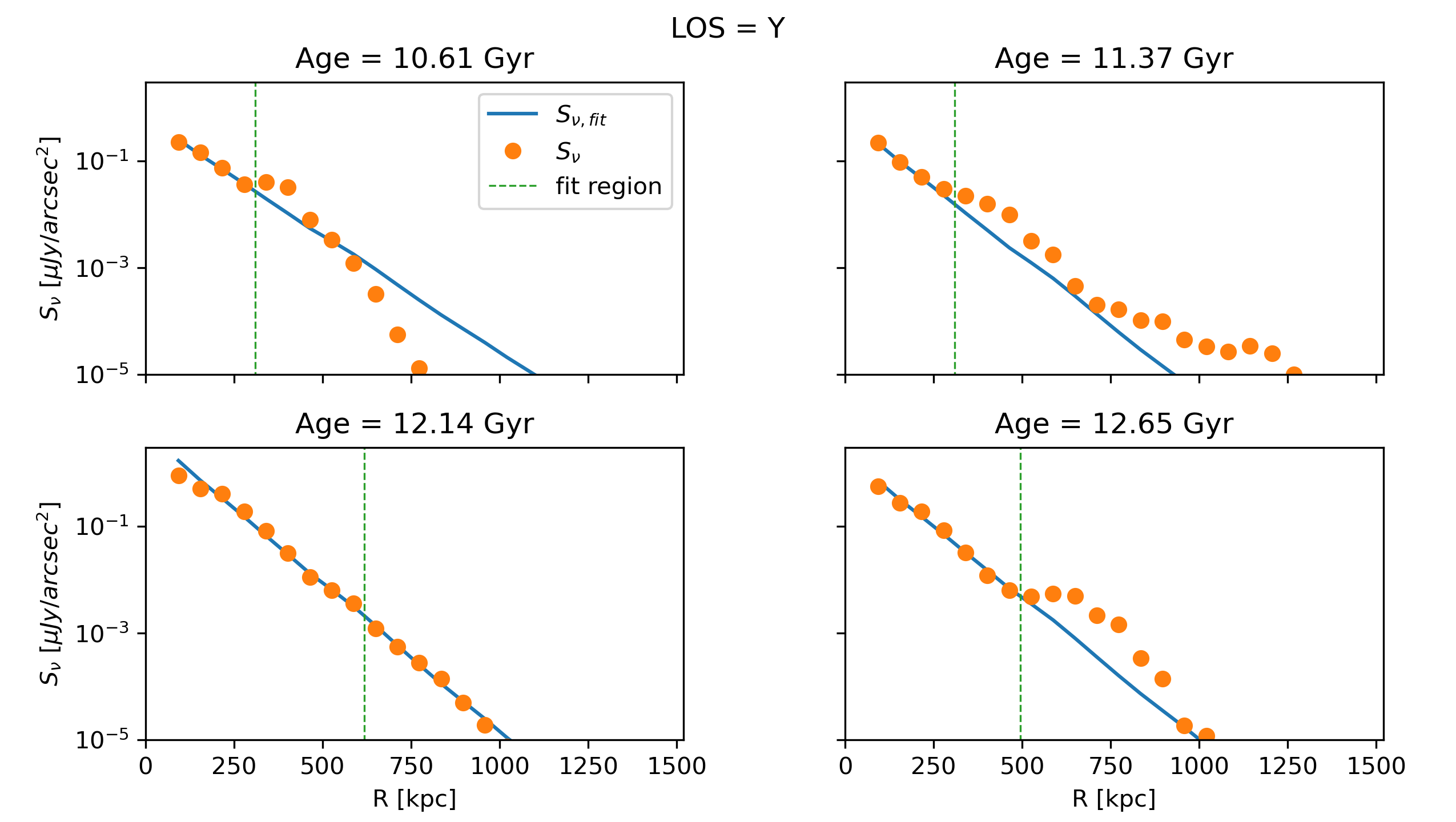}
    \caption{Radial profiles of the cluster at four different epochs along the Y line of sight. The region used for the fit where selected as in \cite{2022Natur.609..911C} and in the figures it is find between the centre and the vertical dashed line.}
    \label{fig:4_rad_prof}
\end{figure*}

\subsection{Evolution of relativistic electrons}

The population of relativistic electrons assigned to the Lagrangian tracers is evolved subject to radiative and collisional losses, adiabiatic terms, and additional (re-)acceleration by shocks and turbulence. We present results for three models ("C","CT" and "CST") already introduced in Sec.~\ref{subsec: crater_roger}.
Model "CST" is the baseline model, which contains all physical losses and plausible (re-)acceleration scenarios expected in the ICM. 
Here, all CRe are injected at 3.37 Gyr ($z\approx2$), with an initial fraction $\phi = 10^{-6}$ with respect the density of thermal electrons, and using $\eta_B = 0.035$ to estimate the unresolved small-scale dynamo on board of tracers. 

The evolving momentum spectra of all CRe in the simulated volume are shown in Fig.~\ref{fig:roger_spt}, where we consider the same snapshots of Fig.~\ref{fig:mass}, with the addition of the initial momentum spectra of all tracers, one time step after the injection of the input power-law distribution.  Early on, when the cluster mass is still composed of several approaching substructures, the CRe spectra mainly suffer Inverse Compton losses.
At 5.96 Gyr ($z \approx 1$) the difference between the three models is not very large. Still, the continuous effect of turbulent re-acceleration has produced a significant excess of high momentum electrons ($\gamma \geq 10^3$). When the additional contribution from merger shocks is included, a small fraction of the CRe population reaches $\gamma \geq 10^5$. 
The actual number of CRe injected by shocks crucially depends here on the assumed acceleration efficiency by DSA, which is highly uncertain \citep[e.g.][]{Bykov19,2023MNRAS.519..548B}. However, for all plausible choices of the acceleration efficiency as a function of Mach number, the bulk of the high-energy population of CRe depends on turbulent re-acceleration acting on the initial seed population and not on the freshly accelerated particles by DSA. This is shown by the small difference between the CT and the CST models in most snapshots. 
In both cases, the plots show a steady growth of the
momentum distribution , namely a large fraction of the momentum distribution has $\gamma \geq 10^3$, through the approximate balance between turbulent re-acceleration, compression and losses, as studied already in Paper I.

Starting from approximately $8 ~\rm Gyr$, recurrent mergers sustain large levels of turbulence within $\rm R_{500}$, and  inject a significant budget of turbulent kinetic energy. By summing all the recorded turbulent kinetic energy by our tracers, we measure up to $2\cdot10^{62}$ erg of solenoidal turbulent kinetic energy injected in the cluster during the assembly of most of the cluster mass, followed by an overall injection of  $3\cdot10^{61}$ erg at the end of the simulation. This turbulent budget, in turn, leads to an increase in CRe energy (both in the CT and CST models), which is key for the formation of extended radio sources. Figure \ref{fig:energy_distribution} shows the ratio between the CRe energy and the kinetic energy measured for all tracers in the MegaRH region, shortly before and after the last important merger experienced by the cluster. The distribution of CRe energy gets significantly wider after the merger, following the re-acceleration by the turbulence injected in this region.
In the $\sim 1.8 \rm ~Gyr$ time difference considered here, the differences in the CRe distribution, or in the ratio between CRe energy and kinetic energy remain significantly larger than the differences of the distribution of turbulent kinetic energy across the same time difference.
This is consistent with the fact (already suggested in our Paper I) that a large fraction of CRe in this regime can reach a long-term balance between gain and losses, while in the same time a significant fraction of the turbulent kinetic energy gets thermalised (consistent with the increase of thermal energy measured over the same timescale). Based on the quantities recorded by tracers, we measure that the solenoidal turbulent kinetic energy is initially $\sim 19\%$  of the total energy after the assembly of the bulk of the cluster mass, and it decreases to $\sim 4\%$ by the end of the simulation, marking a gradual thermalisation of the kinetic energy induced by accretion events. During the same time, however, we measure a steady increase in the estimated total energy of all CRe in our tracers, which amount to $6\cdot10^{-5}\%$ of the total gas energy at $\sim 8 ~\rm Gyr$ and finish having $4\cdot10^{-4}\%$ of the total gas energy towards the end of the simulation.


 It has been shown that the turbulent pressure support relative to the thermal gas energy steadily increases from the cluster centre to the periphery \citep[e.g.][]{va11turbo,2020MNRAS.495..864A}, the turbulent Mach number increases at the cluster periphery and 
the turbulent velocity field becomes increasingly more compressive than solenoidal \citep[e.g.][]{va17turb,2018MNRAS.481L.120V}, and the level of gas clumping and filamentary accretions increase in the outer cluster regions, too \citep[e.g.][]{2021A&A...653A.171A,2022A&A...658A.149S}. Similar results have been obtained by independent groups using different numerical approaches \citep[e.g.][]{in08,lau09,miniati14,2019ApJ...874...42V,2021MNRAS.504..510V}. However, as we have highlighted in Paper I, the radio-emitting particles found in the MegaRH region spent only the last $\sim\rm ~Gyr$ of their evolution there, after being advected to such large radii during the last major merger. Therefore, while the physical properties of turbulence may significantly differ from the innermost to the outermost cluster regions, our simulated electrons experience the full variety of such turbulent fluctuations over their lifetime.  

\begin{figure*}[h!]
    \centering
    \includegraphics[width=0.9\textwidth]{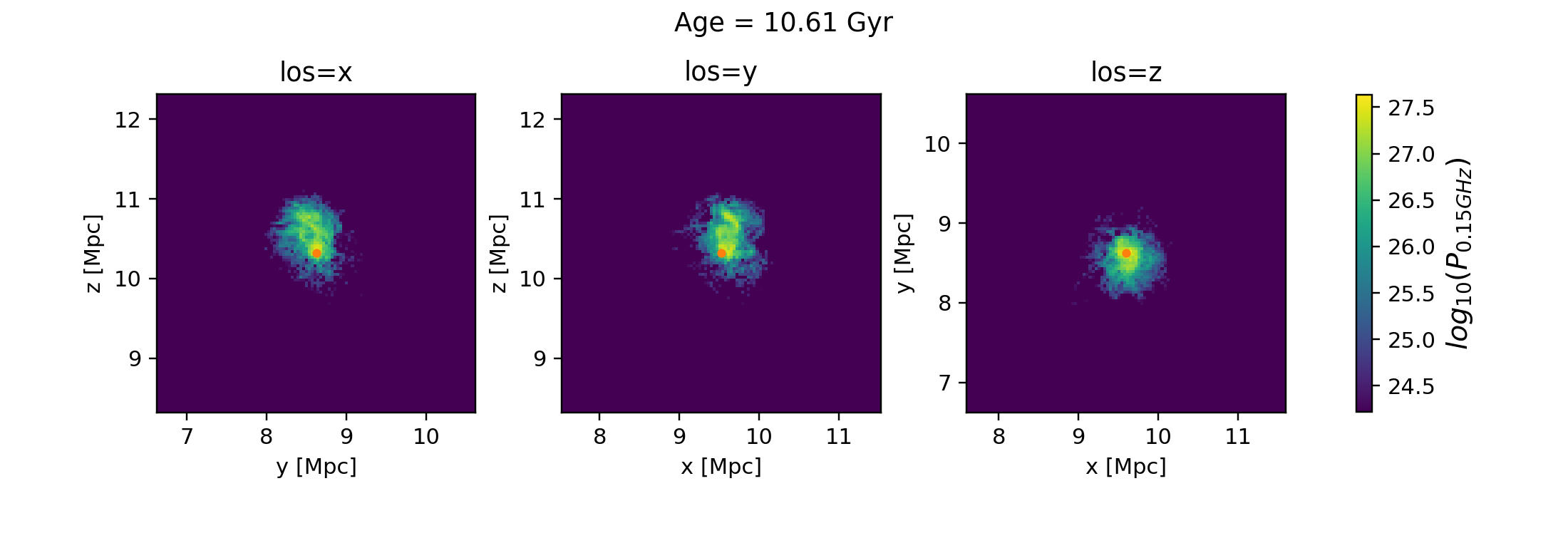}
       \includegraphics[width=0.9\textwidth]{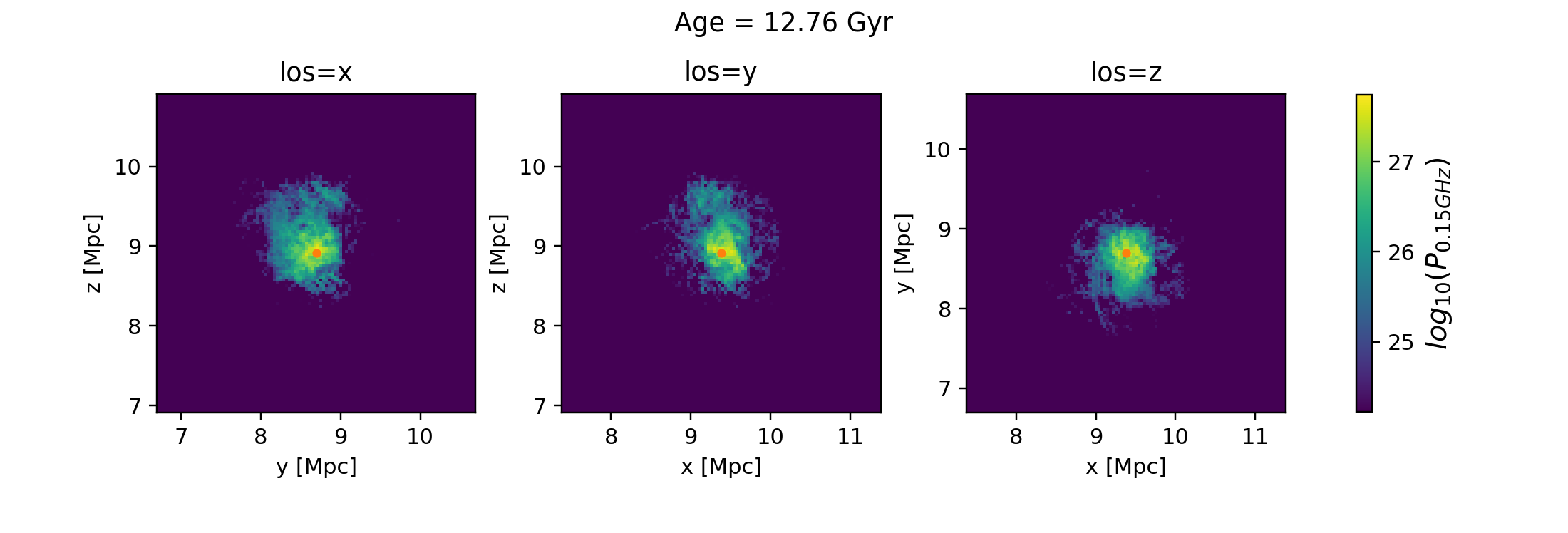}
       \includegraphics[width=0.9\textwidth]{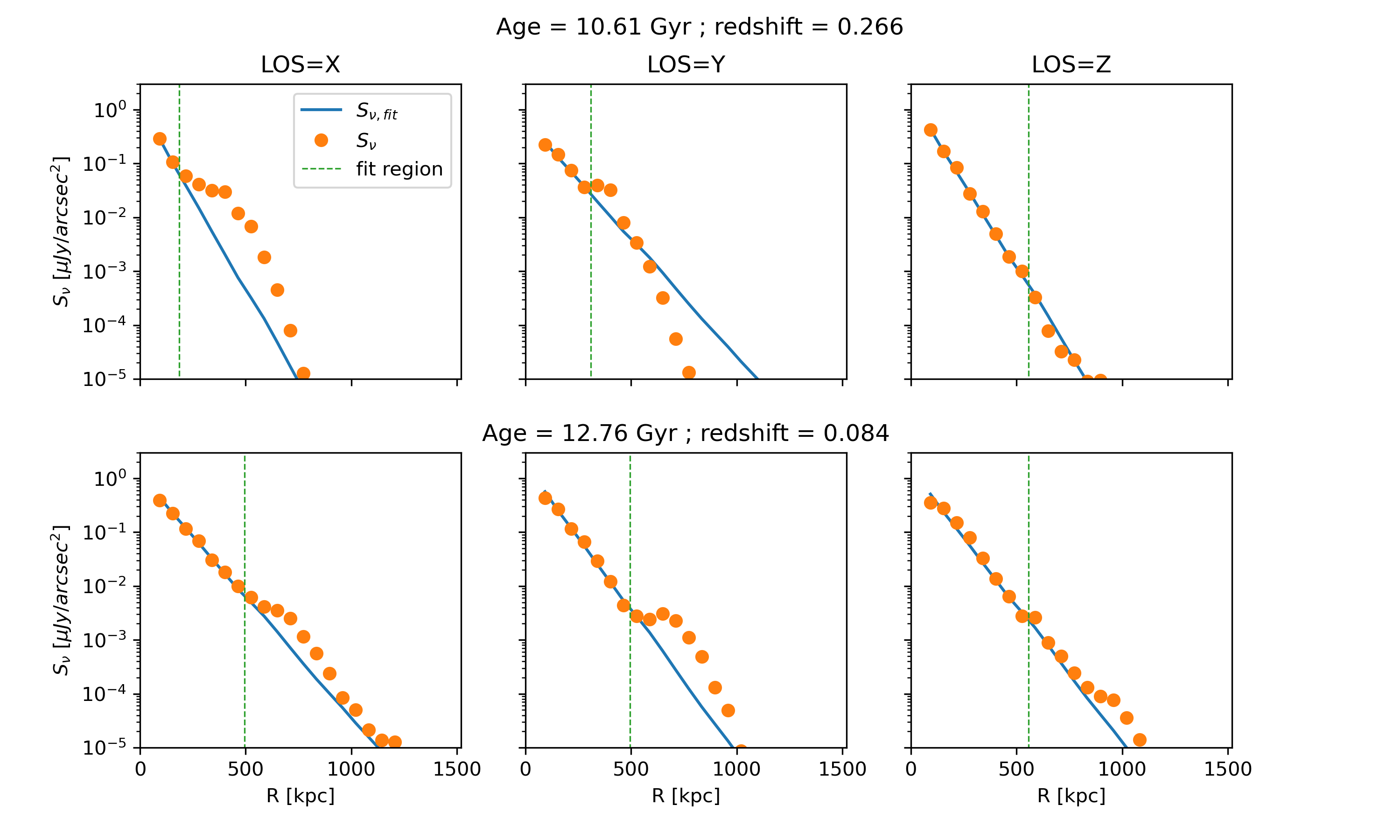}
    \caption{Representations of both the projection of the cluster and its radial profile along that direction at the same ages. Top: Radio maps of the cluster at 10.61 and 12.76 Gyr, along three lines of sight. Bottom: radial profiles of the surface brightness for the three lines of sight. The $z$ line-of-sight roughly corresponds to the merger axis of the latest major merger in the cluster. The region used to compute the best-fit for the radial profile was manually selected as in \cite{2022Natur.609..911C}, and is delimited by the vertical dashed line in the plot.}  \label{fig:projection_effects}
\end{figure*}

\subsection{Evolution of radio emission}

We studied the evolution of the radio properties of the cluster by producing surface brightness map, and their radial profiles, with a procedure similar to \citet{2022Natur.609..911C}.
To this end, we calculate the radio power by summing the synchrotron radio emission of each tracer within a given cell. Then we project the emission along each axis to create mock radio emission maps for the three lines of sight. Finally, we compute the 
radial profiles by radially averaging the surface brightness in concentric annuli, with a width of half of full width at half maximum (FWHM) of the mock observational setup.
Since the exact evolutionary sequence of merger events and radio morphologies experienced by the cluster is randomly associated with redshift (i.e. for slightly different random initial conditions, mergers might have happened with almost equal chance at significantly higher or smaller redshifts) it makes sense to discuss the detectability of the simulated radio emission by fixing a distance between this object and the observer, i.e. to fix a redshift of observation, independent on the actual redshift of each snapshot.
We thus set the cluster distance by fixing a reference redshift of $z_{\rm obs}=0.095$ (which yields the luminosity distance of 449 Mpc), i.e. in the low redshift range of the distribution of potentially detectable MegaRH in the sample of Planck clusters, after using the relation $\rm (1+z)^4\propto M^\beta$ and $\rm \beta = 1$ suggested by \citet{2022Natur.609..911C}, between the host cluster mass ($\rm M_{100}=3.8\times 10^8\ M_{\odot}$) and the minimum redshift for a detection with LOFAR.

While the nominal resolution of LOTSS LOFAR HBA observations is 6" \citep[][]{2019A&A...622A...1S},  in order to enhance the observation's sensitivity to diffuse emission, \cite{2022Natur.609..911C} tapered the LOTSS LOFAR HBA data to a resolution of 60", reaching a rms noise per beam of $\approx 0.2-0.3~ \rm mJy/beam$. Here we follow a similar approach, but consider a slightly smaller tapering length ($39"$, corresponding to 4 resolution elements of the simulation, i.e. $\approx 72 \rm ~ kpc$, at the redshift of our mock observations, $z_{\rm obs}$) because our cluster is smaller than those in \cite{2022Natur.609..911C}. Too much tapering would smooth out features in the radial profile that might distinguish ClassicalRH from MegaRH.

 After computing the surface brightness at this luminosity distance and redshift, we tapered it to a $39" \times 39"$ resolution and compared it to the LOFAR sensitivity to extended radio emission, using as a reference the deepest $\sigma_{\rm rms}=0.2 ~\rm mJy/beam$ sensitivity reached for Abell 2218 in  \citet{2022Natur.609..911C}. Figure~\ref{fig:4snaps} shows the contours of the detectable radio emission at 150 MHz, starting from a $1\sigma_{\rm rms}$ level.
Ignoring any additional contamination by strong point sources and cluster radio galaxies, the diffuse emission would be detectable at the $1 \sigma_{\rm rms}$ level, up to a radius of $\sim 500 ~\rm kpc$ from the cluster centre. This is compatible with a ClassicalRH for a small cluster of galaxies. A more extended component that reaches MegaRH sizes would be too dim to be detected by LOFAR HBA. 

We also compute the radial profile of the surface brightness, by radially averaging the emission within fixed radial annuli from the cluster centre, following a procedure similar to the MegaRH discovery paper by \cite{2022Natur.609..911C}.
The cluster centre is selected as the brightest pixel of the projected image, as is done in observations.
The radial profile of ClassicalRH is
usually taken to follow an exponential profile \citep[e.g.][]{2009A&A...499..679M}:  $\rm I(r) = I_0e^{-r/r_e}$, where $\rm r_e$ is the $e$-folding radius and $\rm I_0$ is the maximum of the luminosity. 
The fit is executed as follows.
First, we performed a preliminary interpolation of the radial surface brightness profile of radio emission with the exponential function, within a manually selected radial range from the peak of radio emission. Although this procedure is semi-empiric and it does not rely on an automatic selection of the radii to associate with a ClassicalRH or a MegaRH region. This is consisent  with the procedure followed in the discovery paper of MegaRH, whose logic we want to reproduce here.
A future larger samples of simulated and observed MegaRH will infer a proper statistics on the extension of this sources, enabling us to design automated procedures for the identification of this structures, 
but this goal goes beyond the purpose of this paper. 
Next, the best fit parameters are used to create a new point symmetric image, by means of the radial profile function. This image is then convolved with a Gaussian beam which $\sigma = FWHM/2.35$ of LOFAR HBA. 
Finally, this image is radially averaged and the best-fit parameters are derived using a least-square method. 
\cite{2022Natur.609..911C} calculated radial profiles by integrating over concentric annuli (including emission below $3\sigma$), and computing the noise as $\rm rms / \sqrt{\rm N_{beams}}$, where $\rm N_{beams}$ is the number of beams within each annulus. The rms is noise calculated from the images where the MegaRH have been detected. 
Considering that our simulated MegaRH would be too dim for detection, we refrain from performing such an analysis here. Our main goal is to assess whether a second, extended component in the surface brightness emerges during the merger, albeit at a low emission level considering the low mass of this system. We also compute the average spectral index between 50 MHz and 150 MHz within the same concentric annuli. 

Figure ~\ref{fig:4snaps} shows the projected radio emission at 150 MHz at four different epochs, along the same line of sight \footnote{A small collection of movies of the radio emission in our cluster at different wavelengths can be seen at \url{https://tinyurl.com/549btyyx}}. This illustrates the variety of the total radio emission produced by the populations of relativistic electrons, and  gives an idea of their detectability.

Figure~\ref{fig:6_maps_b_turb_alpha} shows a longer sequence of evolutionary steps as well as the radial profiles of magnetic fields and turbulent velocities. Also shown is the radial profile of the spectral index of radio emission between 50 and 150 MHz. Already soon after the first major merger leading to the cluster's final mass ($t \sim 8.58 \rm ~Gyr$ in Fig.~\ref{fig:6_maps_b_turb_alpha}, 5a) the cluster forms a centrally located, smooth radio profile, compatible  with a low-power and ultra-steep spectrum radio halo. This source remains detectable for less then 1 Gyr, due to the sudden increase of turbulence following the merger of the parent satellites. This is followed by a phase of low, or absent, radio luminosity for the subsequent 1.5 Gyr, caused by a drop in turbulence and marked by a very steep radio spectral index ($\alpha \leq -1.75$).

The process of turning on and off the diffuse radio emission in a merging cluster is expected to correlate with the cluster crossing time ($\sim R_v/c_s$, where $R_v$ is the cluster virial radius and $c_s$ is the sound speed), as shown in previous studies involving either semi-analytical methods, or numerical simulations of binary mergers \citep[][]{cassano05, 2009A&A...507..661B,2013MNRAS.429.3564D}. However, our simulation allows us to study how the extended radio emission can increase continuously, via the accumulated effect of a series of major or minor mergers, as anticipated in \citep[][]{2023A&A...678L...8B}.

After the first merger, the entrance of a small satellite again boosts the diffuse radio emission above the detection limit. Its effect is observed in the radial profile as a sudden increase of flux in the sector swept by the satellite ($t \sim 10.61 \rm ~Gyr$ in Fig.~\ref{fig:6_maps_b_turb_alpha}, 5c). This event creates a deformation of the profile which does not resemble a MegaRH. However, it still produces significant re-acceleration that enables the formation of extended radio emission at a later stage. Moreover, the launching of merger shock waves at this epoch promotes the formation of a radio relic-like emission feature, powered by shock (re-)acceleration in the northern sector of the cluster. This is also visible in the emergence of a straight power-law in the momentum spectra of Fig.~\ref{fig:roger_spt} 
at 10.7 Gyr, in our CST model which also includes the shock (re-)acceleration of electrons via DSA as explained in Sec.~\ref{sec:methods_sim}, \citep[e.g.][]{inc22,2023MNRAS.526.4234S}. 
The shape of the profile follows the collision geometry (see Fig.~\ref{fig:4_rad_prof}), as can be seen from the asymmetric emission patterns visible in the radio map of Fig.~\ref{fig:4snaps}. 

Starting from $t \sim 11 \rm ~Gyr$, the system enters its most relevant stage for the production of extended emission. This is particularly true for the last major merger, possibly causing the superposition of a ClassicalRH and a MegaRH. During the approaching phase of the merger, the secondary subcluster creates turbulence which re-accelerates particles already in the outskirts of the main cluster, generating a sudden flattening of the radio spectral index at $\rm R_{100}$. The re-energised particles appear first in the north-south direction connecting the main cluster and the accreted subcluster, along a short inter-cluster bridge. This feature may loosely resemble the bridge observed in the Coma cluster \citep[][]{bonafede21}. The bridge remains undetectable with a very low radio luminosity, similar to that of the subcluster itself which itself is not detectable up until the merger.

 The maximum radio luminosity of our cluster is reached at about 12.14 Gyr. 
 When the cluster is observed along a line of sight perpendicular to the merger axis (see e.g. Fig.~\ref{fig:projection_effects}), the radial emission profile appears to be flattened  \citep[this dynamics is reported in a non-cosmological context in][with more detail]{2013MNRAS.429.3564D}. When viewed along the merger axis, we observe an increase in brightness in, both, the central and external region. After $100 \rm ~Myr$ the radio luminosity in the outskirts decreases while the surface brightness of the ClassicalRH remains constant.
 
 After another $100 \rm ~Myr$  (i.e. around 12.65 Gyr in Fig. \ref{fig:4_rad_prof}) we can clearly observe the formation of a second component of the radial profile of surface brightness, which follows from the asymmetric distribution of radio emission boosted by the turbulence in the wake of the accreted subcluster. 
 
 During this epoch, a secondary component in the profile stands out, starting from a radius of $\approx 550 \rm ~kpc$. It remains nearly at the same location for the following $800 \rm ~Myr$. The average surface brightness of this secondary component, reminiscent of a MegaRH, is about two orders of magnitude lower than the peak luminosity of the ClassicalRH, and it remains present in the profile until about $13 ~\rm Gyr$.  The presence of two clear components in the radial profile is seen in the surface brightness, as well as the radio spectrum of the emission. It shows a flattening to about $\alpha \geq -1.5$ in the $550 -1100 \rm ~kpc$ range (compared to the $\alpha \leq -1.7$ of previous epochs), which coincides with the increased level of turbulent velocity at the same radius.

The extended emission in the cluster periphery has some marked similarities with MegaRHs, but also has some important differences: The emerging central diffuse emission in our cluster is mostly compatible in size, power, radial profile and spectral index distribution to ClassicalRHs, which is by itself a key result of our work.  To the best of our knowledge, the only realistic simulation to date of ClassicalRH, using Fokker-Planck methods to study turbulent re-acceleration, were obtained using idealised binary mergers, by \citet[][]{2013MNRAS.429.3564D,2014MNRAS.443.3564D}. Recently, \citet{boss23} simulated the radio emission from the Local cosmic web, including  diffuse radio emission from the Coma cluster. However, from their work and from the distribution of (flat) radio spectra indices it is unclear to which extent the emission is produced by Fermi II acceleration alone and not by DSA acceleration from shocks.

Our simulations clearly show that turbulence injected by the last major merger is important for creating, both, a ClassicalRH and a MegaRH-like source.
Two closely related effects contribute to this: 
the merger generates a significant turbulent wake along the merger axis, with rms velocities comparable to those injected in the cluster centre, and the large-scale motions induced by the merger displace a large pool of relativistic electrons that are already re-energised by past acceleration cycles. After 800 Myr, such re-energised electrons get dispersed to more peripheral cluster regions, i.e. as far as $R_{\rm 200}$, while being still barely detectable in radio. In both cases, these mechanisms tend to produce a rather asymmetric emission towards the external cluster regions, which is at variance with about a half of known MegaRH \citep[][]{2022Natur.609..911C}.
However, the actual luminosity changes with time as well as with the direction of the observation.

\begin{figure*}[h!]
    \centering
    \sidecaption
    \includegraphics[width=12cm]{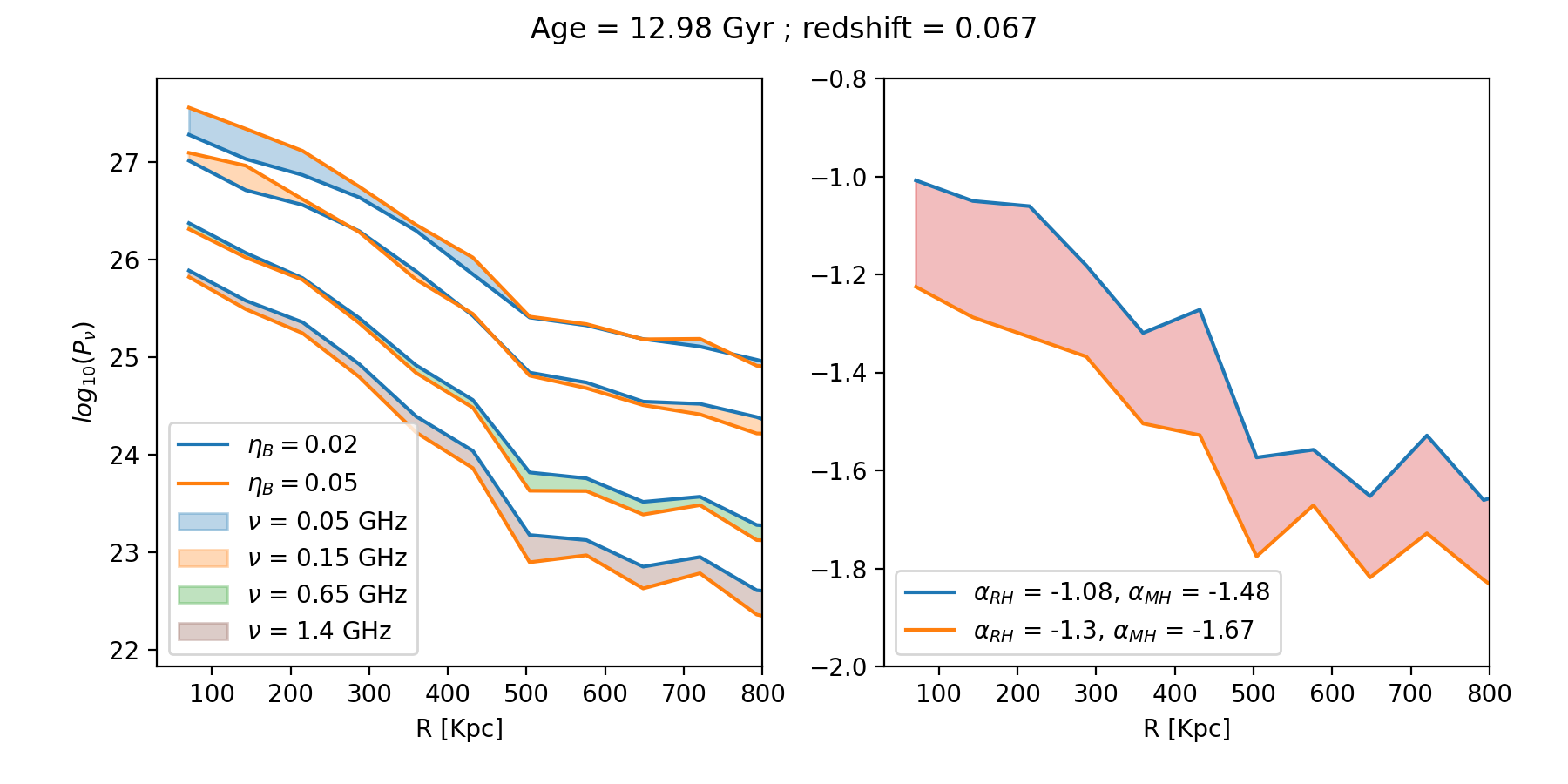}
    \caption{Radial profile of the radio power with all the frequencies and spectral index radial distribution. Both plots show the profiles computed with two different magnetic field strengths: one with $\eta_B=0.02$ (blue line) and the other with $\eta_B=0.05$ (orange line). The filled region represents all the intermediate values that the magnetic field can assume. The right panel also shows the values of the spectral index in the two regions of MegaRH and ClassicalRH.}  \label{fig:mag_field}
\end{figure*}

\subsection{Exploring the formation of mega radio halos} 

Under the obvious caveat that this pilot study is based on a single  cluster of galaxies that undergoes a series of mergers, our results shed  important light on how very extended radio emission (including Classical and MegaRH-like structures) forms. 

In the simulation, the fundamental mechanism responsible for the formation of diffuse radio emission, both, in the form of Classical and Mega RHs, is CR re-acceleration by turbulence. This turbulence is injected on scales of $\sim \rm Mpc$ during mergers. This turbulence is more solenoidal and more subsonic in the ClassicalRH regions, than in the MegaRH, where it is more compressive and has a higher Mach number.
Indeed based on the gas temperature profile in this region at $\sim 12.65 ~\rm Gyr$(see also Paper I), the local sound speed is $c_s\sim 960 \rm ~km/s$ on average in the Classical RH region, while it is  $c_s\sim 750 \rm ~km/s$ in the MegaRH region;  at the same time,  the rms velocity related to the solenoidal component is  $\sigma_v  \sim 280~ \rm km/s$ in the ClassicalRH region and $\sigma_v \sim 250~ \rm km/s$ in the MegaRH region, corresponding to a turbulent Mach number $\mathcal{M}_t \sim 0.29$ in the first case, and to $\mathcal{M}_t \sim 0.33$ in the second case.
 During the post-merger phase, we observe a significant re-energisation of electrons. This results in the appearance of a second component in the radial brightness profiles, as is observed in real  MegaRH emission. Re-energisation is primarily triggered by the significant increase in turbulent velocity and a comparatively slower rise in the magnetic field strength.
The radio emitting electrons of the MegaRH were located in the cluster centre before the last merger, and got advected to outer radii (and encountered significantly different turbulent conditions) only their last 
  $\sim \rm ~Gyr$ of evolution \citep[][]{2023A&A...678L...8B}. This process allows a large fraction of CRe to keep a $\gamma \gtrsim 10^3$ energy level, which makes them subject to stochastic turbulent re-acceleration for timescales longer than their cooling time (consistent with what anticipated in Paper I). 

The formation of a second component in the radial profile of surface brightness, as well as a change in the spectral index of the emission, are in general not observable along all possible lines of sight through the cluster because the MegaRH-like structure produced in our model is not spherically symmetric, but rather ellipsoidal.

As an example, in Fig.~\ref{fig:projection_effects} shows the Classical and of the MegaRH regions at times of 10.61 and 12.76 Gyr, viewed along the three coordinate axes. 

The main axis of the emission is roughly aligned with the direction of the last major merger, roughly aligned with the $z$-axis of the simulation. This marks the direction where a fraction of re-accelerated electrons are spread from the innermost cluster regions to the cluster outskirts. Therefore, the most striking appearance of a MegaRH emission pattern are found at a perpendicular angle along the merger axis (i.e. along axis $x$ and $y$), i.e. when the last important merger in the cluster happens close to the plane of the sky. 
Conversely, if the cluster is observed along the direction parallel to the merger axis ($z$), the mock observation detects a mostly ClassicalRH exponential profile of surface brightness.

We stress that under realistic observing conditions, our simulated MegaRH would hardly be detectable. Its low mass makes the whole process subcritical to produce radio emission beyond the LOFAR HBA and LBA detection limit, if we place the system at $z \approx 0.1$. 
In principle, the results of this study can be qualitatively extrapolated to the regime of higher-mass clusters, considering the approximate self-similar dependence between the turbulent kinetic energy and the cluster mass, i.e. $E_{\rm turb} \propto M^{5/3}$ \citep[e.g.][]{cassano05,2006MNRAS.369L..14V}, and the approximate scaling between the radio luminosity and the kinetic power, i.e. $L_R \propto E_{\rm turb}/t_{\rm eddy}$, where $t_{\rm eddy} \sim L/\sigma_v$ is the eddy turnover time for an injection scale L \citep[in turn, typically a fraction of the cluster virial radius, e.g.][]{bv20} and $\sigma_v$ is the rms velocity dispersion computed within the same scale.
However, other processes are likely not to obey similar self-similar rescaling relations, e.g. the injection scale of turbulence, the damping of MHD waves as well as the ratio between synchrotron losses and Inverse Compton losses. For this reason, it is presently uncertain how to extrapolate these results to clusters with 10 times larger mass, which are closer to the real observations. This will be subject of forthcoming work.

\begin{figure*}[h!]
    \centering
    \sidecaption
    \includegraphics[width=12cm]{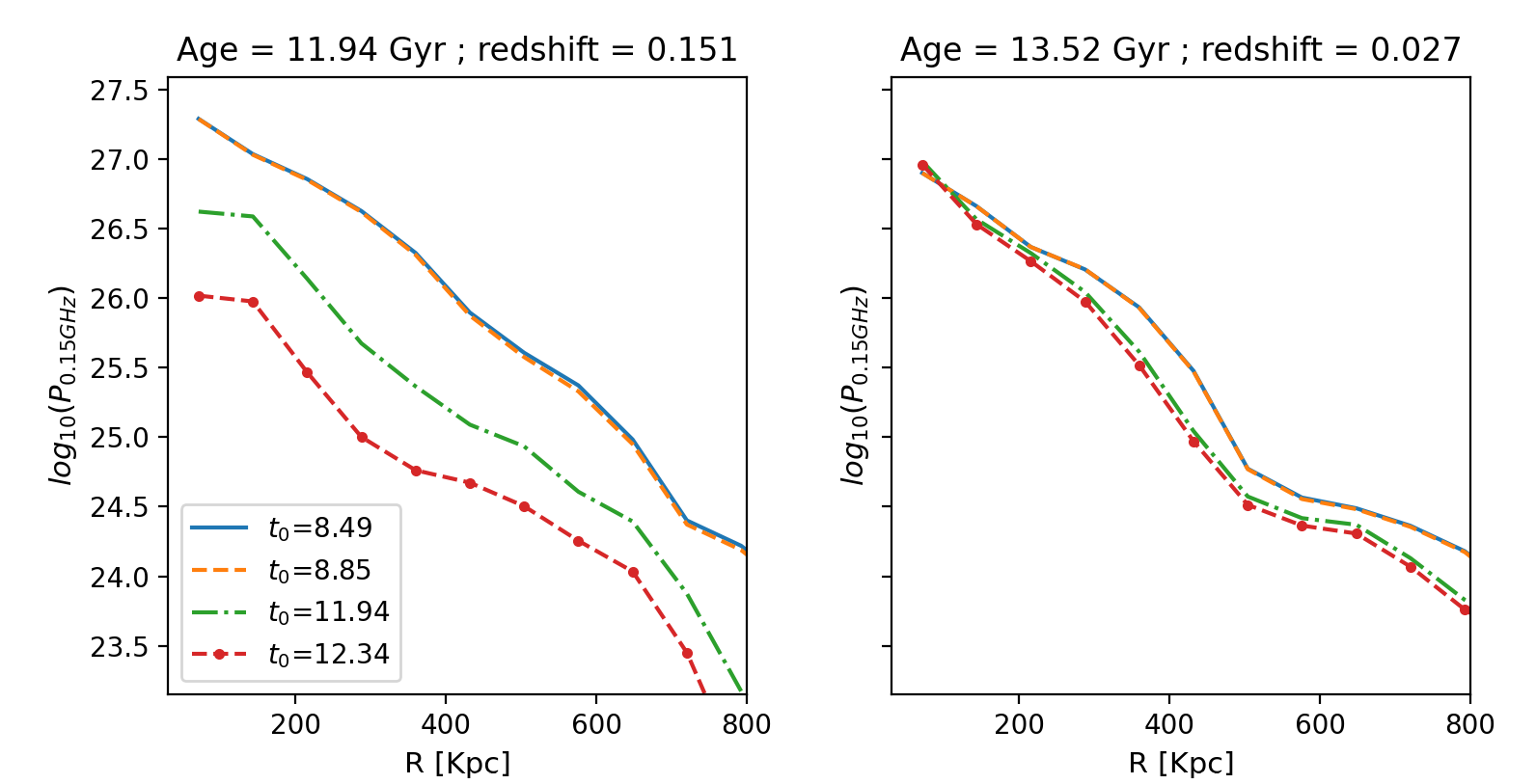}
    \caption{Radial profiles of radio surface brightness at 150MHz, considering  four different injection times for the seed CRe. The injection epochs were chosen to bracket the important moments marking the assembly of most of the cluster mass, at $t \approx 8.6 \rm ~Gyr$, as well as the last major merger experienced by the cluster, at $t \approx 12 \rm Gyr$ 
    (see also Fig.\ref{fig:mass}).} 
    \label{fig:t_inj}
\end{figure*}

\subsection{Model variations} 
\label{roger_setting}

We tested our baseline model under a few important variations of numerical and physical parameters. 
In particular, we have run ROGER with different initial conditions, i.e. by varying the fraction of relativistic electrons to thermal electrons, by modifying the assumed efficiency of the small-scale dynamo, and by initialising the initial electron population at different times.  
By comparing the radial emission profiles and the spectra of radio emission at the end of the simulation, we found that the only parameter responsible for large variations of the radio emission is the initial fraction of relativistic electrons assigned to the tracer particles ($\phi=10^{-6}$ in the baseline model). Changing this fraction linearly affects the resulting radio emission. 
While the latter result is trivial for the case of re-acceleration solely acting on fossil relativistic electrons, in our baseline CST model also the injection of new (i.e. freshly accelerated) relativistic electrons is allowed in our model. Even though the injection of new electrons by merger shocks can locally enhance the population of electrons and their radio emission (producing radio relic-like emission features), the number of CRe accumulated in our cluster is dominated by the initial population.\\

The variations in the assumed magnetic field amplification efficiency are more subtle to address, as an increase in the magnetic field has multiple effects on the evolution of radio emitting electrons: a higher magnetic field causes a faster cooling of electrons (but limited to magnetic fields exceeding the CMB equivalent, otherwise Inverse Compton losses remain the leading cooling mechanism). It also increases the synchrotron emissivity, but it also reduces the efficiency of Fermi-II acceleration because in the assumed turbulent re-acceleration model $\tau_{\rm acc} \propto |B|$ \citep[][]{bv20}. 
As we can observe in Fig.~\ref{fig:mag_field}, varying the assumed amplification efficiency of the magnetic field from $\eta_B=0.02$ to $\eta_B=0.05$ (while the value used in the baseline model is $\eta_B=0.035$)  does not cause significant variation in the cluster radio luminosity. The largest effect is found in the central region for low frequencies, whereas the effect is small in the outer regions. 

Interestingly, at higher frequencies (1.4 GHz) we observe the opposite trend. A weaker magnetic field enables electrons to reach higher energies in the outskirts, while showing no significant difference in the central luminosity.  Moreover, the variation in the magnetic field produces a small shift in the radial distribution of the radio spectral index.\\

Finally, we varied the epoch for the first injection of the seed electron population associated with our tracers.
In detail, we resimulated the evolution of radio emissions for tracers, considering injection just before or after the assembly of most of the cluster mass (which is roughly located at a cosmic time $\sim 8.6 \rm ~Gyr$ since), as well as just before or after the last major merger experienced by the cluster (cosmic time $\sim 12 \rm ~Gyr$).
These variations have overall  a negligible effect, unless in the unrealistic case in which all electrons are injected just in the latest 2 Gyr of the simulation, i.e. after the last major merger. 

In Fig.~\ref{fig:t_inj} we can see that noticeable differences are found only when (implausibly) late injection of electrons is considered, while otherwise the radio emission predicted in the last evolutionary stages of the cluster is largely unaffected by variations in the initial population.
This result implies that, under the continuous effect of turbulent re-acceleration, electrons injected at all epochs maintain high energies for most of the time. This is sufficient to keep particles at an approximate balance between losses and gains.

\section{Discussion and conclusions} 
\label{sec:conclusions}

We have presented a new framework to use cosmological simulations for the modelling of diffuse radio emission in clusters of galaxies. 
We analysed the evolution of relativistic electrons subject to energy losses and gains. We used Lagrangian tracer particles to track the propagation of families of relativistic electrons injected in the simulation since $z=2$ \citep[as in][]{2023A&A...678L...8B} and we ran a 
parallel Fokker-Planck algorithm \citep[based on][]{2013MNRAS.429.3564D} to model the energy evolution across the $1 \leq P/(m_e c) \leq 10^6$ range of CRe momenta.

In a first cosmological simulations, we report the emergence of diffuse radio emission caused by turbulent re-acceleration of fossil relativistic electrons injected several $\sim $Gyrs before the assembly of the cluster.
In general, we find that the turbulence produced by mergers is  sufficient to re-accelerate already relativistic electrons via Fermi-II processes, in the adiabatic-stochastic-acceleration scenario \citep[][]{2016MNRAS.458.2584B,bv20}.

We produced mock radio images of the simulation and we analysed them similarly to \cite{2022Natur.609..911C}. 
The radial emission profile of the cluster shows that mergers are capable of injecting enough turbulent kinetic power to re-accelerate relativistic electrons and make them emit detectable radio emission on large scales. When the innermost cluster regions become radio-bright, they produce diffuse emission structures similar to ClassicalRH, in terms of their radial profile and their radio spectra.
Moreover, the asymmetric and elongated diffuse radio emission that develops in our simulated cluster, after the latest major mergers, can reasonably  reproduce some key features of real MegaRH, even if the observed sample is still small, and our single simulation does not allow us to derive very firm conclusions. The extended and peripheral emission structures formed in our simulation have very steep spectra, which make them very difficult to be detected beyond $\geq 120 \rm ~MHz$, which appears also in line with available data. 
With the exploration of several variations of our baseline re-acceleration model (i.e. by varying the epoch of injection of fossil electrons, their normalisation as well as the amplitude of magnetic field amplification) we verified that this picture is relatively insensitive to the unknown plasma parameters of the re-acceleration model. 

In our simulation, MegaRH develop in regions where turbulence has a higher turbulent Mach number than in the innermost regions. Transsonic turbulence decays more quickly \citep[see][on decaying turbulence in the ICM]{2019MNRAS.488.3439S}, which affects the spectra of velocities and magnetic fields. Hence these regions also produce weaker radio emission, with a typically steep radio spectrum. 

However, particle (re)-acceleration in transsonic conditions may favour TTD over ASA, since the latter is more efficient for turbulent Mach numbers in the range $\mathcal{M_{\rm turb}}<0.2-0.5$ turbulence \citep[e.g.][]{2016MNRAS.458.2584B,bv20}.
Unlike for ASA, in TTD there is an extra free parameter to set, namely the dissipation efficiency onto magnetosonic waves. Testing different combinations of TTD/ASA and their prediction on spectra and statistics of MegaRH is an exciting avenue for further work.

Another important aspect which can only be explored with a larger sample of simulated clusters, is the role played by the merger mass ratio. While it is at most 1:5 in the cluster studied in this first pilot study, 
smaller mass ratio can lead to an increased relative amount of turbulent injection. For example, \citet{cassano05} estimated with a semi-analytical treatment of ram-pressure stripping that a 1:2 merger induces about twice as a much turbulent energy (relative to the thermal energy), than a 1:5 merger, which would implies a more efficient acceleration of electrons.
However, it is difficult to predict also the associated effect of shock dissipation and shock injection of electrons on the formation of diffuse radio emission, and direct and time-dependent simulations are best suited to study such non-linear process in detail, especially in the little explored regime of MegaRH.
In conclusion, our study supports the idea that diffuse radio emission in clusters of galaxies can be produced by the turbulent re-acceleration of relativistic electrons.  In particular, both ClassicalRH and MegaRH form following major mergers and the resulting dissipation of solenoidal kinetic energy. MegaRH result from a more anisotropic diffuse radio component that is elongated approximately along the direction of the last major merger. 
They also result from more solenoidal and more supersonic turbulence.

The relativistic electrons contributing the most to the diffuse radio emission are fossil electrons injected at high redshifts ($z \geq 2$). These fossil electrons are injected into the highest gas density peaks in the ratio of $10^{-6}$ with respect to the thermal electron density. This is compatible with the level of CRe seeding expected from the cumulative activity of radio galaxies and star formation winds \citep[e.g.][and references therein]{2024arXiv240316068V}.

 While we could simulate only a low-mass cluster  ($M_{100} \sim 3 \cdot 10^{14}$ at $z=0$), in future work we will validate our model against simulations of clusters whose masses match those for which MegaRHs have been detected. Moreover, we will account for the effect of a time-dependent injection of fossil electrons by active galactic nuclei.

\section*{Acknowledgements}
We thank our anonymous reviewer for the precious suggestions and comments to the first version of this paper.
F.V. has been supported by Fondazione Cariplo and Fondazione CDP, through grant n. Rif: 2022-2088 CUP J33C22004310003 for "BREAKTHRU" project. F.V. also  acknowledges the usage of computing  at the Gauss Centre for Supercomputing e.V. (www.gauss-centre.eu) for supporting this project by providing computing time through the John von Neumann Institute for Computing (NIC) on the GCS Supercomputer JUWELS at J\"ulich Supercomputing Centre (JSC), under projects "BREAKTHRU". FV acknowledges the usage of computing time on the LEONARDO HPC System at CINECA, under project "IsB28\_RADGALEO", with F.V. as Principal Investigator. 
MB acknowledges funding by the Deutsche Forschungsgemeinschaft (DFG, German Research Foundation) under Germany's Excellence Strategy -- EXC 2121 ``Quantum Universe'' --  390833306 and project number 443220636 (DFG research unit FOR 5195: "Relativistic Jets in Active Galaxies").

\bibliographystyle{aa}
\bibliography{franco3}

\appendix

\section{Fokker-Planck method and loss terms}
\label{appendix1}
We closely followed the method outlined by \citet{2013MNRAS.429.3564D}, which we incorporated in the existing version of the ROGER code \citep[][]{va23a}, to solve the evolution of the number density of relativistic electrons in momentum space, $N(p)$, using the following Fokker-Planck equation: 

\begin{align} 
\label{fp2}
    \frac{\partial N(p)}{\partial t} &= \frac{\partial}{\partial p} \left[ H(p) N(p)  + D_{\mathrm{pp}} \frac{\partial N(p)}{\partial p} \right] \nonumber\\ 
    & + Q_{\mathrm{inj}}(p,t)
\end{align}

where $Q_{\mathrm{inj}}(p,t)$ is a time-dependent source term for the injection of electrons (in our case, only allowed for DSA) and 
$H(p)$ is the generalised cooling function $H(p)$: 

\begin{align}
    H(p) &= -\frac{2}{p}D_{\mathrm{pp}} + \sum_i \frac{dp}{dt}_{i} ,
\end{align}
which combines all relevant losses for the particle population:  ionisation and Coulomb losses, adiabatic losses, synchrotron and Inverse Compton losses,  as well as the gain by turbulent acceleration, via the $D_{\mathrm{pp}}$ term.

While the acceleration terms have been discussed in the main paper (Sec. 2.2), the loss terms considered in our work are:
\begin{itemize}
\item  Coulomb losses, expressed as:
\begin{equation}
    \left(\frac{dp}{dt}\right)_\mathrm{c} = -3.3 \cdot 10^{-29} n_\mathrm{th}\left[1+\frac{ln(\gamma/n_\mathrm{th})}{75}\right]\ ,
\end{equation}
where $n_\mathrm{th}$ is the thermal proton density in the medium in $\rm cm^{-3}$ and $\gamma$ is the Lorentz factor. 
\item Synchrotron and Inverse Compton losses, expressed as:
\begin{equation}
    \left(\frac{dp}{dt}\right)_\mathrm{rad} = -4.8 \cdot 10^{-4} p^2\left[\left(\frac{B_\mathrm{\mu G}}{3.2}\right)^2+(1+z)^4\right]\ ,
\end{equation}
where, $z$ is the redshift, $B_\mathrm{\mu G}$ is the ICM magnetic field in $\mu \mathrm{G}$ and where we assumed isotropic distribution of momenta with respect to the magnetic field. The term between the square brackets represents the magnetic field strength of the ICM, while the second is the equivalent magnetic field strength that an electron would experience by interacting with the background of CMB photons during an IC scattering process.

\item reversible adiabatic changes (either losses or gains) are modelled through the divergence of the velocity field, which is measured on board of tracers:

\begin{equation}
\left(\frac{dp}{dt}\right)_\mathrm{ad} = \frac{2 ~p}{3} \cdot (\vec{\nabla} \cdot \vec{v}) \ ,
\end{equation}

\end{itemize}

We neglect bremsstrahlung losses since their timescale is significantly larger than the ones of all other loss channels for the ICM physical condition considered here. 

We solved the  equations based on the  finite difference scheme by \citet{1970JCoPh...6....1C} and  with the method by \citet{2013MNRAS.429.3564D}, which incorporates both systematic and stochastic effects of Fermi II acceleration, together with all most important cooling effects.  This scheme ensures particle number conservation by construction, it is suitable for logarithmic grids and it is unconditionally stable, allowing to us to use a small number of momentum bins. 

\end{document}